\documentclass[]{aastex631}

\usepackage{amsmath}

\definecolor{ForestGreen}{RGB}{34,139,34}
\shorttitle{}
\shortauthors{Gaurav N. Gadbail et al.}
\graphicspath{{./}{figures/}}

\begin{document}

\title{Gaussian Process Approach for Model-Independent Reconstruction of $f(Q)$ Gravity with Direct Hubble Measurements}

\author[0000-0003-0684-9702]{Gaurav N. Gadbail}
\affiliation{Department of Mathematics, Birla Institute of Technology and
Science-Pilani,\\ Hyderabad Campus, Hyderabad-500078, India.}

\author[0000-0003-2570-2335]{Sanjay Mandal}
\affiliation{Faculty of Symbiotic Systems Science,\\
 Fukushima University, Fukushima 960-1296, Japan.}

\author[0000-0003-2130-8832]{P.K. Sahoo}\thanks{pksahoo@hyderabad.bits-pilani.ac.in}
\affiliation{Department of Mathematics, Birla Institute of Technology and
Science-Pilani,\\ Hyderabad Campus, Hyderabad-500078, India.}



\begin{abstract}
 The increase of discrepancy in the standard procedure to choose the arbitrary functional form of the Lagrangian $f(Q)$ motivates us to solve this issue in modified theories of gravity. In this regard, we investigate the Gaussian process (GP), which allows us to eliminate this issue in a $f(Q)$ model-independent way. In particular, we use the 58 Hubble measurements coming from cosmic chronometers and the radial Baryon acoustic oscillations (BAO) to reconstruct $H(z)$ and its derivatives $H'(z)$, $H''(z)$, which resulting lead us to reconstruct region of $f(Q)$, without any assumptions. The obtained mean curve along $\Lambda$CDM constant in the reconstructed region follows a quadratic behavior. This motivates us to propose a new $f(Q)$ parametrization, i.e., $f(Q)= -2\Lambda+ \epsilon Q^2$, with the single parameter $\epsilon$, which signifies the deviations from $\Lambda$CDM cosmology. Further, we probe the widely studied power-law and exponential $f(Q)$ models against the reconstructed region and can improve the parameter spaces significantly compared with observational analysis. In addition, the direct Hubble measurements, along with the reconstructed $f(Q)$ function, allow the $H_0$ tension to be alleviated.

\end{abstract}



\section{Introduction}
In contemporary cosmology, the adoption of modified gravity theories has proven highly efficacious in explaining the late-time and near-time acceleration of the Universe \citep{Capozziello_2011,Nojiri_2011}, circumventing the need for postulating dark energy or inflationary components \citep{Peebles_2003,Cai_2010,Li_2019,Li_2019_2,Elizalde_2019}. Recently, with the mounting discrepancy in the Hubble constant ($H_0$), researchers have become increasingly inclined to investigate modified gravity as a means to resolve cosmological tension \citep{Wong_2019,Freedman_2019,Vagnozzi_2020}. This motivation arises from the inadequacy of the $\Lambda$CDM (Lambda Cold Dark Matter) model in addressing these tensions \citep{Cai_2016}. Numerous modified gravity theories have been proposed in the literature, with a recent emphasis on a theory rooted in non-metric scalar $Q$, which is solely geometric in nature \citep{Jim_nez_2018}. This modified gravity theory is developed under the assumptions of torsionlessness and a vanishing Ricci scalar.

In contemporary discourse, this modified theory of gravity is in high demand due to its successful portrayal of various cosmological scenarios. Extensive research has been conducted within this framework to address current cosmological issues \citep{Lazkoz_2019,Anagnostopoulos_2021,Frusciante_2021,Lin_2021,Mandal_2020,Mandal_2020_2,Jim_nez_2020,Harko_2018,heisenberg2023review}. Additionally, several modifications or extensions of this theory, such as $f(Q,T)$ gravity \citep{Xu_2019} and $f(Q,B)$ or $f(Q,C)$ gravity \citep{Capozziello_2023,De_2024}, have been proposed. However, a fundamental challenge associated with modified gravity theories lies in the arbitrary selection of a function governing the nonmetricity scalar $Q$. This approach often necessitates making numerous assumptions and lacks inherent symmetry.

Furthermore, observational data are employed to constrain the cosmological models involving $f(Q)$ and to estimate the requisite parameters for desired outcomes \citep{Lazkoz_2019,Anagnostopoulos_2021,Ayuso_2021}. Various methodologies exist for discussing the cosmophysical properties, including assuming specific forms of $f(Q)$, studying the dynamical behavior of backgrounds and perturbations, and validating outcomes through observational testing, including comparisons with the solar system\citep{Jim_nez_2020,Khyllep_2021,Khyllep_2023}.

While these methodologies provide valuable insights, advancements in learning have facilitated a process whereby the function $f(Q)$ can be constructed using observational measurements without presupposing a particular form. This reconstruction process, known as the Gaussian Process (GP) method, has been successfully developed and utilized across various studies \citep{Holsclaw_2010,exp1,exp2}. This method has been studied and explored in various dark energy scenarios (see the references herein \citep{Melia_2018,Pinho_2018,Zhang_2018,Yin_2019,Elizalde_2019,Rau_2019, G_mez_Valent_2018,Fortunato_2024}) and the expansion history of the universe \citep{Busti_2014,verde2014expansion,Li_2016,wang2016modelindependent,Melia_2018,Cai_2020N,Ren_2022N}. This procedure allows for the reconstruction of $f(Q)$ in a model-independent manner, thereby enhancing the robustness of cosmological analyses. Consequently, the GP process is poised to play a pivotal role in modern cosmological inquiries, enabling the representation of reconstructions in terms of uncertainty and offering a means to reconstruct $f(Q)$ without assuming specific conditions.

The investigation in this study unfolds across several sequential stages. Initially, we provide a succinct overview of the symmetric teleparallel gravity framework for the FLRW spacetime metric in Section \ref{section 2}, succeeded by an exploration of Gaussian processes with a focus on reconstructing the Hubble parameter and its derivative in Section \ref{section 3}. Moving to Section \ref{section 4}, we meticulously outline the step-by-step procedures employed in the reconstruction process of \( f(Q) \), also confronted particular selections for \( f(Q) \) against it. Additionally, we delve into cosmological applications to corroborate the prevailing state of the universe. Finally, in Section \ref{section 5}, we encapsulate our findings and contemplate future perspectives.

\section{$f(Q)$ Gravity Theory}
\label{section 2}
We begin this section by discussing the metric affine connection - a fundamental mathematical tool in differential geometry and general relativity, providing a framework for understanding the geometry of curved manifolds equipped with both metric and affine structures. To investigate the cosmological aspects of non-metric gravity, let us examine the most general form of the affine connections
\begin{equation}
\hat{\Gamma}^{\,\sigma}_{\,\,\,\mu\nu}=\Gamma^{\,\sigma}_{\,\,\,\mu\nu}+K^{\,\sigma}_{\,\,\,\mu\nu}+L^{\,\sigma}_{\,\,\,\mu\nu},
\end{equation}
where the Levi-Civita connection $\Gamma^{\,\sigma}_{\,\,\,\mu\nu}$ is defined as
\begin{equation}
\Gamma^{\,\sigma}_{\,\,\,\mu\nu} =\frac{1}{2}g^{\sigma\lambda}\left(\partial_{\mu}g_{\lambda\nu}+\partial_{\nu}g_{\lambda\mu}-\partial_{\lambda}g_{\mu\nu}\right),
\end{equation}
which can be uniquely determined by the first-order derivatives of the metric tensor $g_{\mu\nu}$. The contortion $K^{\,\sigma}_{\,\,\,\mu\nu}$ and deformation tensor $L^{\,\sigma}_{\,\,\,\mu\nu}$ are defined as
\begin{eqnarray*}
K^{\,\sigma}_{\,\,\,\mu\nu} &=& \frac{1}{2}T^{\,\sigma}_{\,\,\,\mu\nu}+T^{\,\,\,\,\,\,\,\sigma}_{(\mu\,\,\,\,\,\,\nu)},\\
L^{\,\sigma}_{\,\,\,\mu\nu} &=& -\frac{1}{2}g^{\sigma\lambda}\left(Q_{\mu\lambda\nu}+Q_{\nu\lambda\mu}-Q_{\lambda\mu\nu}\right),
\end{eqnarray*}
respectively, which describes non-Riemannian properties in the manifold. The contortion tensor disappears in the symmetric teleparallel theory because it follows an anti-symmetric property. The interplay between nonmetricity and the absence of torsion would influence cosmological models and the evolution of the universe. These effects could manifest in scenarios such as the dynamics of inflation, the behavior of dark energy, and the formation of large-scale structures.\\
The non-metricity tensor $Q_{\sigma\mu\nu}$ is defined as
\begin{equation}
\label{1}
Q_{\sigma\mu\nu}=\nabla_{\sigma}g_{\mu\nu},
\end{equation}
and the corresponding traces are $ Q_{\sigma}=Q_{\sigma\,\,\,\,\mu}^{\,\,\,\,\mu}$, $\tilde{Q}_{\sigma}=Q^{\mu}_{\,\,\,\,\sigma\mu}$.
Aside from that, the superpotential tensor $P_{\,\,\mu\nu}^{\sigma}$ can be written as 
\begin{equation}
\label{3}
4P_{\,\,\mu\nu}^{\sigma}=-Q^{\sigma}_{\,\,\,\,\mu\nu}+2Q^{\,\,\,\,\,\,\sigma}_{(\mu\,\,\,\,\nu)}-Q^{\sigma}g_{\mu\nu}-\tilde{Q}^{\sigma}g_{\mu\nu}-\delta^{\sigma}_{(\mu}\, Q\,_{\nu)},
\end{equation} 
obtaining a trace of nonmetricity tensor or nonmetricity scalar as 
\begin{equation}
\label{4}
Q=-Q_{\sigma\mu\nu}P^{\sigma\mu\nu}.
\end{equation}  
In this work, we study the extension of symmetric teleparallel theory called $f(Q)$ gravity theory, and its considered action is given as  \citep{Jim_nez_2018}
\begin{equation}
\label{5}
S=\int \left\{\frac{1}{2\kappa^2}\left[Q+f(Q)\right]+\mathcal{L}_m\right\}\sqrt{-g}\,d^4x,
\end{equation}
where $f(Q)$ represents any function of the scalar $Q$, $g$ denotes the determinant of $g_{\mu\nu}$, and $\mathcal{L}_m$ stands for the matter Lagrangian density.\\
As action \eqref{5} varies with respect to the metric, the gravitational field equation for $f(Q)$ is obtained, and it is written as
\begin{equation}
\label{7}
\frac{2}{\sqrt{-g}}\nabla_{\sigma}\left((1+f_Q)\sqrt{-g}\,P^{\sigma}_{\,\,\mu\nu}\right)+\frac{1}{2}(Q+f(Q))\,g_{\mu\nu}+(1+f_Q)\left(P_{\mu\sigma\lambda}Q_{\nu}^{\,\,\,\sigma\lambda}-2Q_{\sigma\lambda\mu}P^{\sigma\lambda}_{\,\,\,\,\,\,\nu}\right)=- T_{\mu\nu},
\end{equation}
where $f_Q=\frac{d f}{d Q}$. The energy-momentum tensor for matter is now defined as
$T_{\mu\nu}\equiv-\frac{2}{\sqrt{-g}}\frac{\delta(\sqrt{-g})\mathcal{L}_m} {\delta g^{\mu\nu}}$.\\
To utilize $f(Q)$ gravity in a cosmological context, we adopt the spatially flat Friedmann-Lemaitre-Robertson-Walker (FLRW) spacetime, characterized by a specific metric
\begin{equation}
\label{10}
ds^2=-dt^2+a^2(t)\,\delta_{ij}\,dx^i\,dx^j,\,\,\,\,\,\,(i,j=1,2,3),
\end{equation}
corresponding nonmetricity scalar is obtained as $Q=6H^2$, where $H=\frac{\dot{a}}{a}$ is the Hubble parameter with $a(t)$ denoting cosmological scale factor and the upper dot denotes derivative with respect to the coordinate time $t$. Applying the FLRW metric into the general field equation \eqref{7}, the relevant Friedman equations of $f(Q)$ cosmology, namely
 
\begin{equation}
\label{11}
H^2+2H^2\,f_Q-\frac{f}{6}=\frac{\rho_m}{3},
\end{equation} 
\begin{equation}
\label{12}
\left(12H^2\,f_{QQ}+f_Q+1\right)\dot{H}=-\frac{1}{2}(p_m+\rho_m),
\end{equation}
where $f_Q=\frac{df}{dQ}$, and $f_{QQ}=\frac{d^2f}{dQ^2}$. Furthermore, in the equations provided, $\rho_m$ represents the energy density, and $p_m$ denotes the pressure of the matter fluid. It can be easily derived that they accomplish the conservation equation $\dot{\rho_m}+3H(\rho_m+p_m)=0$. \\
We can rewrite Eqs. \eqref{11} and \eqref{12} as the standard form
\begin{equation}
\label{13}
    3H^2=\rho_m+\rho_{DE},
\end{equation}
\begin{equation}
\label{14}
    2\dot{H}+3H^2=-(p_m+p_{DE}),
\end{equation}
where 
\begin{equation}
\label{rhode}
    \rho_{DE}=\frac{f}{2}-Q\,f_Q,
\end{equation}
\begin{equation}
\label{pde}
    p_{DE}=2\dot{H}\left(2Q\,f_{QQ}+f_Q\right)-\rho_{DE},
\end{equation}
are the dark energy density and pressure contributed by the modified part of geometry. Then, by using Eqs. \eqref{rhode} and \eqref{pde}, we can define the effective dark energy equation of state as
\begin{equation}
\label{w}
    \omega_{DE}=\frac{p_{DE}}{\rho_{DE}}=-1+\frac{4\dot{H}\left(2Q\,f_{QQ}+f_Q\right)}{f-2Q\,f_Q}.
\end{equation}
Additionally, the conservation equation of the effective dark energy,
\begin{equation}
\dot{\rho}_{DE}+3H(\rho_{DE}+p_{DE})=0.
\label{c}
\end{equation}
In our analysis, we focus on the late-time evolution of the cosmic fluid, so that we can neglect radiation and consider the entire contribution due to pressureless matter. This implies $p_m=0$ and $\rho_m=3H_0^2\,\Omega_{0m}(1+z)^3$, where the subscript zero refers to quantities evaluated at the present time, and $z$ is the redshift defined as $z=\frac{1}{a}-1$.

\section{Gaussian Processes Using Observational Hubble Data}
\label{section 3}
\subsection{Gaussian Process}
 A Gaussian process is a type of statistical model that extends a Gaussian distribution. Gaussian process regression is a technique that is commonly used to reconstruct functions and their derivatives directly from observed data without making any assumptions. This process involves gathering a set of random variables that all follow a Gaussian distribution \citep{Seikel_2012}. The relationship between these variables is determined by a covariance matrix function, which is uniquely determined by the data points. As a result, Gaussian processes provide a way to reconstruct functions without relying on any specific physical assumptions or parameterizations.\\
The Gaussian process is written as \citep{Seikel_2012,Mehrabi_2021,exp1}
\begin{equation}
    f(x) \mathtt{\sim} \mathcal{GP}\left(\mu(x),k(x,\Tilde{x})\right)
\end{equation}
where $k(x,\Tilde{x})=\mathbb{E}[(f(x)-\mu(x))(f(\Tilde{x})-\mu(\Tilde{x}))]$ is the kernel function and $x$ are the observational points. The $\mu(x)=\mathbb{E}[f(x)]$ provides the mean of the random variable at each $x$. In this work, we employ the squared exponential function as our kernel function to reconstruct functions and their derivatives \citep{exp1,Mehrabi_2021,exp2}. This kernel function represents the most versatile form of covariance function, and it is given by

\begin{equation}
    k(x,\Tilde{x})=\sigma^2_f\,exp\left(-\frac{(x-\Tilde{x})^2}{2\,l^2}\right)
\end{equation}
This kernel function depends on the two hyperparameters $\sigma_f$ and $l$. Specifically, $l$ determines the correlation length between consecutive values of $f(x)$, while $\sigma_f$ regulates the variation of $f(x)$ in relation to the process mean.\\
In this study, we utilize the Gaussian Processes in Python (GAPP) developed by Seikel et al. \citep{Seikel_2012}, to reconstruct the evolution of the Hubble function \(H(z)\) and its derivatives from observational Hubble data.

\subsection{Observational Hubble Data (OHD)}
\begin{figure*}
\gridline{\fig{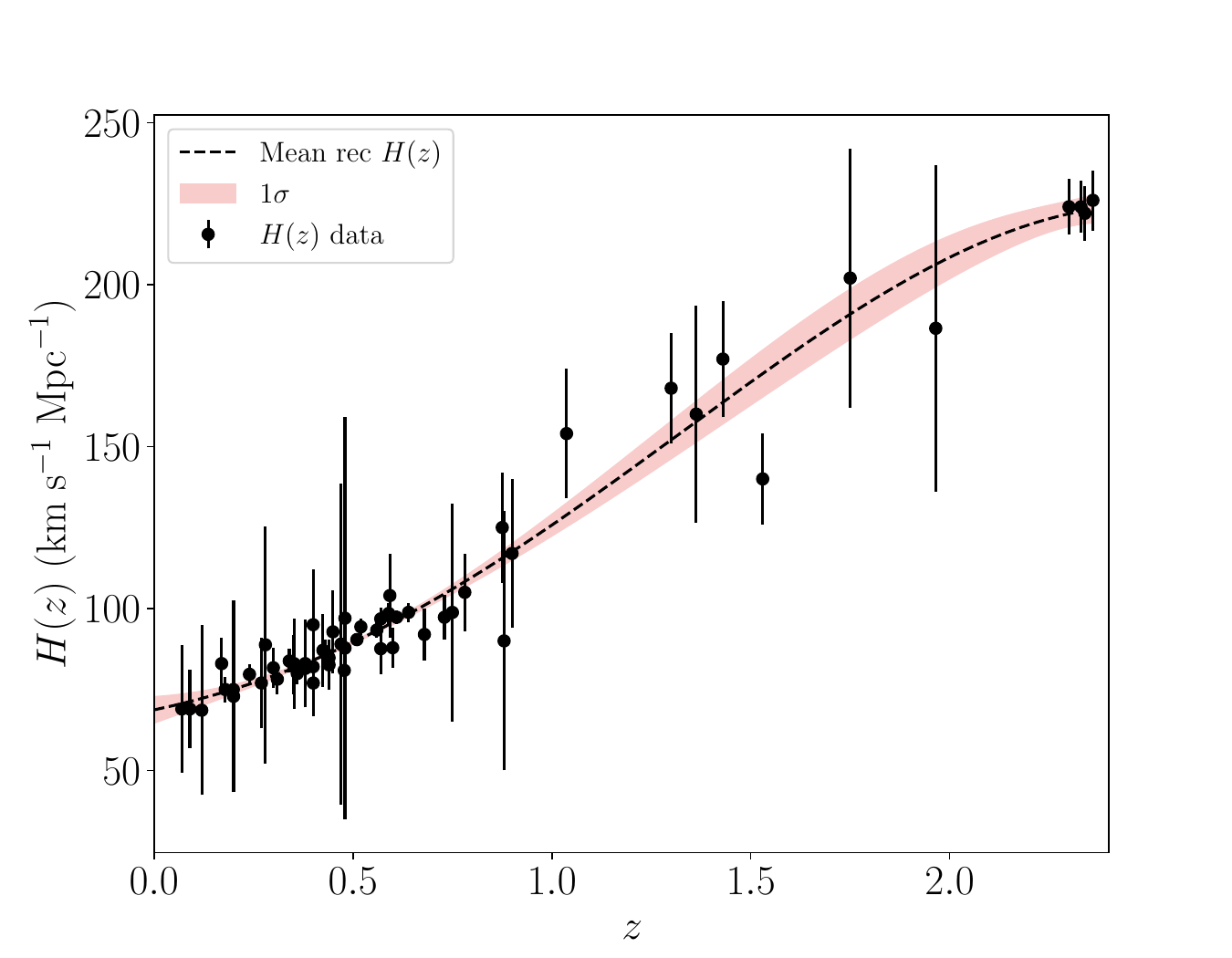}{0.5\textwidth}{(a)}
          \fig{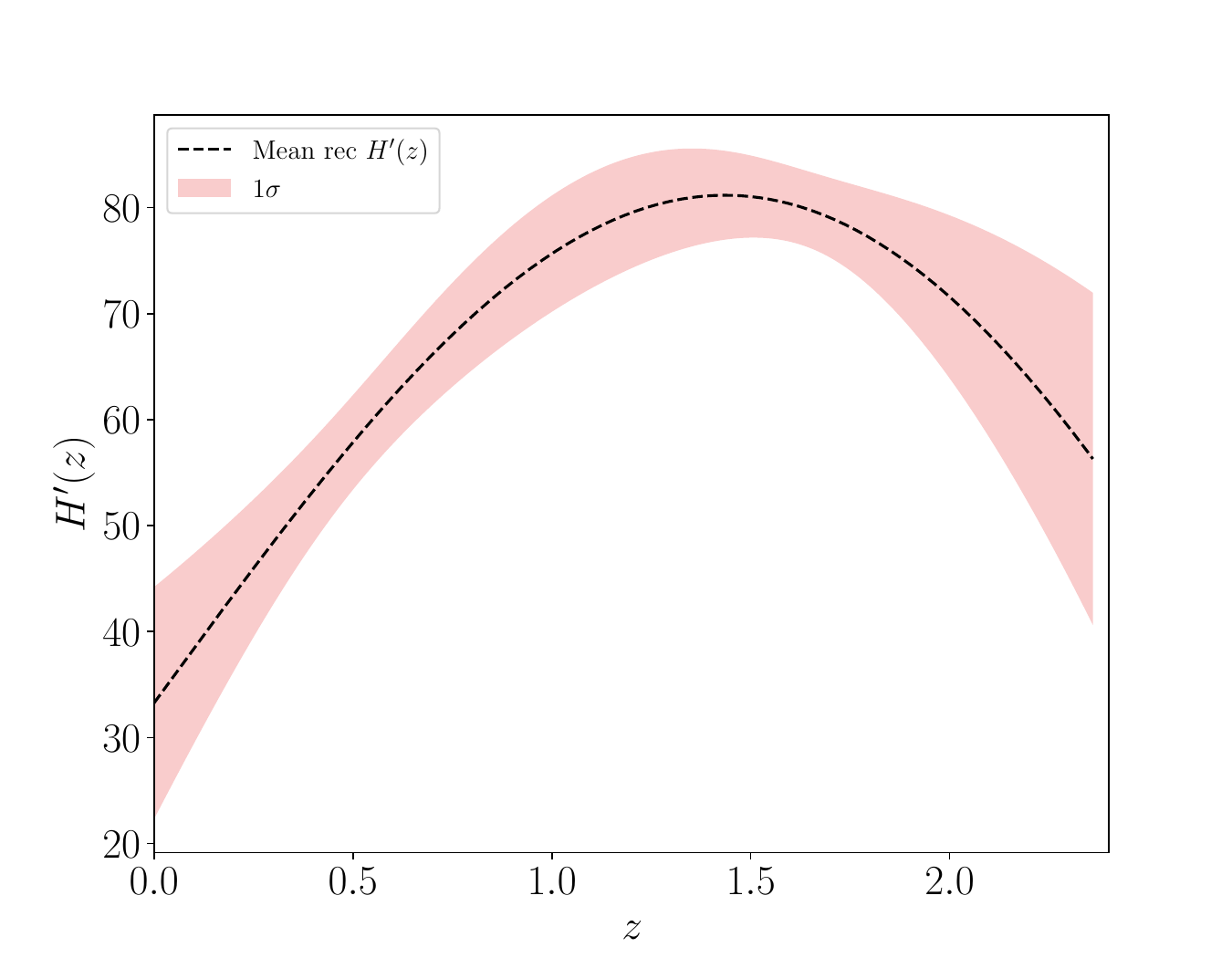}{0.5\textwidth}{(b)}
          }
\caption{In the left panel, we can see the reconstructed behavior of $H(z)$, which is derived from the 32 CC data points and the 26 BAO data points of the radial method. In the right panel, we can see the reconstructed behavior of the derivative of $H(z)$ with respect to $z$. The black dashed line in each graph represents the mean reconstructed curve, while the colored region indicates $1\sigma$ errors arising from the GP errors.  \label{Hz}}
\end{figure*}

We used the latest 58 points of Hubble data along with their error bars for the Gaussian reconstruction process. Out of these 58 points, 32 were obtained from cosmic chronometer (CC) observations, which provide information on $H(z)$ from the age evolution of passively evolving galaxies in a model-independent way. The remaining 26 points were obtained from radial baryon acoustic oscillation (BAO) observations which measure the clustering of galaxies with the BAO peak position as a standard ruler.  The BAO peak position depends on the sound horizon. BAO data are model-dependent. But, if we use only CC data, no useful constraint could be found for $f(z)$; therefore, we use the combination of two Hubble samples, which increases the statistics and helps us find better GP results. The OHD comprises 58 data points within the redshift range of $0.07<z<2.42$. From this scrutiny, we determined the $H_0$ value as $H_0=68.71\pm4.3$ $km\,s^{-1}\,\,Mpc^{-1}$. We have presented a figure comparing with the recent measurements on $H_0$ in Figure \ref{H0}. In Table \ref{Table 1}, we present the above points along with references. As we are considering the direct and local measurements of the Hubble values; the $H_0$ tension will be alleviated by the modified $f(Q)$ gravity reconstruction.\\
The Hubble parameter \(H(z)\) and its derivative \(H'(z)\) (prime denote the derivative with respect to $z$), successfully reconstructed in a model-independent manner, are depicted in Figure \ref{Hz}.

\begin{table}[]


        \caption{Here, table contains the $58$ points of Hubble parameter values $H(z)$ with errors $\sigma _{H}$ from differential age ($32$ points), and BAO and other ($26$ points) approaches, along with references.}
        \label{Table 1}     
\begin{tabular}{|c c c c c c c c|}\hline
\multicolumn{8}{|c|}{Table-1: $H(z)$ datasets consisting of 58 data points} \\ \hline
\multicolumn{8}{|c|}{CC data (32 points)}  \\ \hline
$z$ & $H(z)$ & $\sigma _{H}$ & Ref. & $z$ & $H(z)$ & $\sigma _{H}$ & Ref. \\ \hline
$0.070$ & $69$ & $19.6$ & \cite{Daniel_2010} & $0.4783$ & $80.9$ & $9$ & \cite{Moresco_2016} \\ \hline
$0.09$ & $69$ & $12$ & \cite{Simon_2005} & $0.480$ & $97$ & $62$ & \cite{Daniel_2010} \\ \hline
$0.120$ & $68.6$ & $26.2$ & \cite{Daniel_2010} & $0.593$ & $104$ & $13$ & \cite{M_Moresco_2012} \\ \hline
$0.170$ & $83$ & $8$ & \cite{Simon_2005} & $0.6797$ & $92$ & $8$ & \cite{M_Moresco_2012} \\ \hline
$0.1791$ & $75$ & $4$ & \cite{M_Moresco_2012} & $0.75$  & $98.8$  & $33.6$ & \cite{Borghi_2022} \\ \hline
$0.1993$ & $75$ & $5$ & \cite{M_Moresco_2012} & $0.7812$ & $105$ & $12$ & \cite{M_Moresco_2012} \\ \hline
$0.200$ & $72.9$ & $29.6$ & \cite{Zhang_2014} & $0.8754$ & $125$ & $17$ & \cite{M_Moresco_2012} \\ \hline
$0.270$ & $77$ & $14$ & \cite{Simon_2005} & $0.880$ & $90$ & $40$ & \cite{Daniel_2010} \\ \hline 
$0.280$ & $88.8$ & $36.6$ & \cite{Zhang_2014} & $0.900$ & $117$ & $23$ & \cite{Simon_2005} \\ \hline 
$0.3519$ & $83$ & $14$ & \cite{M_Moresco_2012} & $1.037$ & $154$ & $20$ & \cite{M_Moresco_2012} \\ \hline 
$0.3802$ & $83$ & $13.5$ & \cite{Moresco_2016} & $1.300$ & $168$ & $17$ & \cite{Simon_2005} \\ \hline 
$0.400$ & $95$ & $17$ & \cite{Simon_2005} & $1.363$ & $160$ & $33.6$ & \cite{Moresco_2015} \\ \hline 
$0.4004$ & $77$ & $10.2$ & \cite{Moresco_2016} & $1.430$ & $177$ & $18$ & \cite{Simon_2005} \\ \hline
$0.4247$ & $87.1$ & $11.2$ & \cite{Moresco_2016} & $1.530$ & $140$ & $14$ & \cite{Simon_2005} \\ \hline
$0.4497$ & $92.8$ & $12.9$ & \cite{Moresco_2016} & $1.750$ & $202$ & $40$ & \cite{Simon_2005} \\ \hline
$0.470$ & $89$ & $49.6$ & \cite{Ratsimbazafy_2017} & $1.965$ & $186.5$ & $50.4$ & \cite{Moresco_2015}   \\ \hline
\multicolumn{8}{|c|}{From BAO \& other method (26 points)} \\ \hline
$z$ & $H(z)$ & $\sigma _{H}$ & Ref. & $z$ & $H(z)$ & $\sigma _{H}$ & Ref. \\ \hline
$0.24$ & $79.69$ & $2.99$ & \cite{Gaztaga_2009} & $0.52$ & $94.35$ & $2.64$ & \cite{Wang_2017} \\ \hline
$0.30$& $81.7$ & $6.22$ & \cite{Oka_2014} & $0.56$ & $93.34$ & $2.3$ & \cite{Wang_2017} \\ \hline
$0.31$ & $78.18$ & $4.74$ & \cite{Wang_2017} & $0.57$ & $87.6$ & $7.8$ & \cite{Chuang_2013} \\ \hline
$0.34$ & $83.8$ & $3.66$ & \cite{Gaztaga_2009} & $0.57$ & $96.8$ & $3.4$ & \cite{Anderson_2014} \\ \hline
$0.35$ & $82.7$ & $9.1$ & \cite{Chuang_2013} & $0.59$ & $98.48$ & $3.18$ & \cite{Wang_2017} \\ \hline
$0.36$ & $79.94$ & $3.38$ & \cite{Wang_2017} & $0.60$ & $87.9$ & $6.1$ & \cite{Blake_2012} \\ \hline
$0.38$ & $81.5$ & $1.9$ & \cite{Alam_2017} & $0.61$ & $97.3$ & $2.1$ & \cite{Alam_2017} \\ \hline
$ 0.40$ & $82.04$ & $2.03$ & \cite{Wang_2017} & $0.64$ & $98.82$ & $2.98$ & \cite{Wang_2017}  \\ \hline
$0.43$ & $86.45$ & $3.97$ & \cite{Gaztaga_2009} & $0.73$ & $97.3$ & $7.0$ & \cite{Blake_2012} \\ \hline
$0.44$ & $82.6$ & $7.8$ & \cite{Blake_2012} & $2.30$ & $224$ & $8.6$ & \cite{Busca_2013} \\ \hline
$0.44$ & $84.81$ & $1.83$ & \cite{Wang_2017} & $2.33$ & $224$ & $8$ & \cite{Bautista_2017} \\ \hline
$0.48$ & $87.79$ & $2.03$ & \cite{Wang_2017} & $2.34$ & $222$ & $8.5$ & \cite{Delubac_2015} \\ \hline
$0.51$ & $90.4$ & $1.9$ & \cite{Alam_2017} & $2.36$ & $226$ & $9.3$ & \cite{Font_Ribera_2014} \\ \hline
\end{tabular}
 \end{table}

\section{Reconstructing the $f(Q)$ Function from Gaussian Processes Utilizing OHD Data}
\label{section 4}
In this section, we will attempt to derive the functional form of $f(Q)$ by using the reconstructed Hubble function and its derivative, which we obtained in the previous section by applying the Gaussian process to OHD data. The process of reconstruction is simpler in FLRW cosmology in $f(Q)$ gravity, as it only depends on the Hubble function and its first-order derivative. Our goal is to establish the relationship between the redshift $z$ and $f$, or in other words, to find $f(z)$.\\
To use the model-independent reconstruction approach, we need first to extract the expressions for the involved derivatives $f_Q$ as
\begin{equation}
    f_Q\equiv \frac{df}{dQ}=\frac{df/dz}{dQ/dz}=\frac{f'}{12HH'}
\end{equation}
where primes represent the derivative with respect to redshift $z$. The following step in the application of the GP is to take the approximation of $f'$ as
\begin{equation}
\label{20}
    f'(z)\approx \frac{f(z+\Delta z)-f(z)}{\Delta z},
\end{equation}
for small $\Delta z$.
Let’s compute the approximation error. We write a Taylor expansion of $f(z + \Delta z)$ about $z$, and then we obtain
\begin{equation}
\label{21}
    f'(z)=\frac{f(z + \Delta z)-f(z)}{\Delta z}-\frac{\Delta z}{2}f''(\zeta),\,\,\,\,\,\zeta\in (z,z+\Delta z).
\end{equation}
The second term on the right-hand side of \eqref{21} is the error term. Since the approximation \eqref{20} can be thought of as being obtained by truncating this term from the exact formula \eqref{21}, this error is called the truncation error.
The small parameter $\Delta z$ denotes the distance between the two points $z$ and $z+\Delta z$. As this distance tends to zero, i.e., $\Delta z\to 0$, the two points approach each other, and we expect the approximation \eqref{20} to improve. This is indeed the case if the truncation error goes to zero, which in turn is the case if $f''(\zeta)$ is well defined in the interval $(z,z+\Delta z)$. The “speed” in which the error goes to zero as $\Delta z \to 0$ is called the rate of convergence. When $\Delta z\to 0$, it could increase the compatibility of the reconstructed model with the $\Lambda$CDM model.\\
Using the modified Friedmann equation \eqref{f1} and the approximation above for $f'(z)$, we can extract a recursive relation between consecutive redshifts ($z_i$ and $z_{i+1}$). This involves writing $f(z_i+1)$ as a function of $f(z_i)$, and $H(z_i)$ and $H'(z_i)$ as
\begin{equation}
    f(z_{i+1})-f(z_i)\\=-6(z_{i+1}-z_i)\frac{H'(z_i)}{H(z_i)}\left[H^2(z_i)-\frac{f(z_i)}{6}-\frac{\rho_m(z_i)}{3}\right].
\end{equation}
Ultimately, we arrived at the final phrase as follows by using the EoS parameter for the matter sector:
\begin{equation}
    f(z_{i+1})=f(z_i)-6(z_{i+1}-z_i)\frac{H'(z_i)}{H(z_i)}\\\left[H^2(z_i)-\frac{f(z_i)}{6}-H_0^2\,\Omega_{m0}(1+z_i)^3\right].
\end{equation}
Utilizing the provided expression, we can compute the value of $f$ at the redshift $z_{i+1}$, given that we possess information about the parameters at redshift $ z_i$. Furthermore, through an analysis of the connection between $Q$ and $H$, and by observing the evolution of $H(z)$, we can derive the expression of $f$ in relation to redshift $z$.

\begin{figure*}
\gridline{\fig{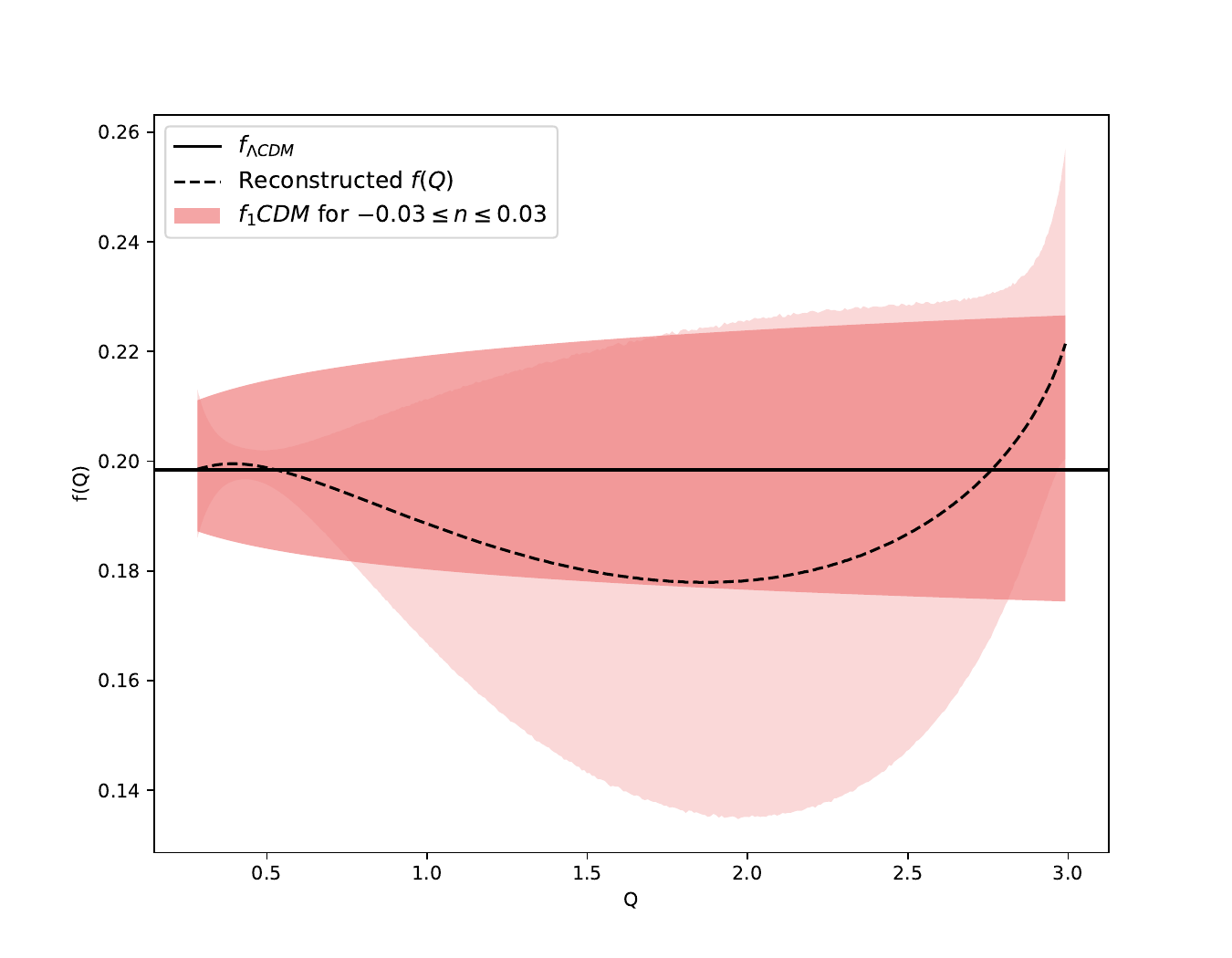}{0.5\textwidth}{(a)}
          \fig{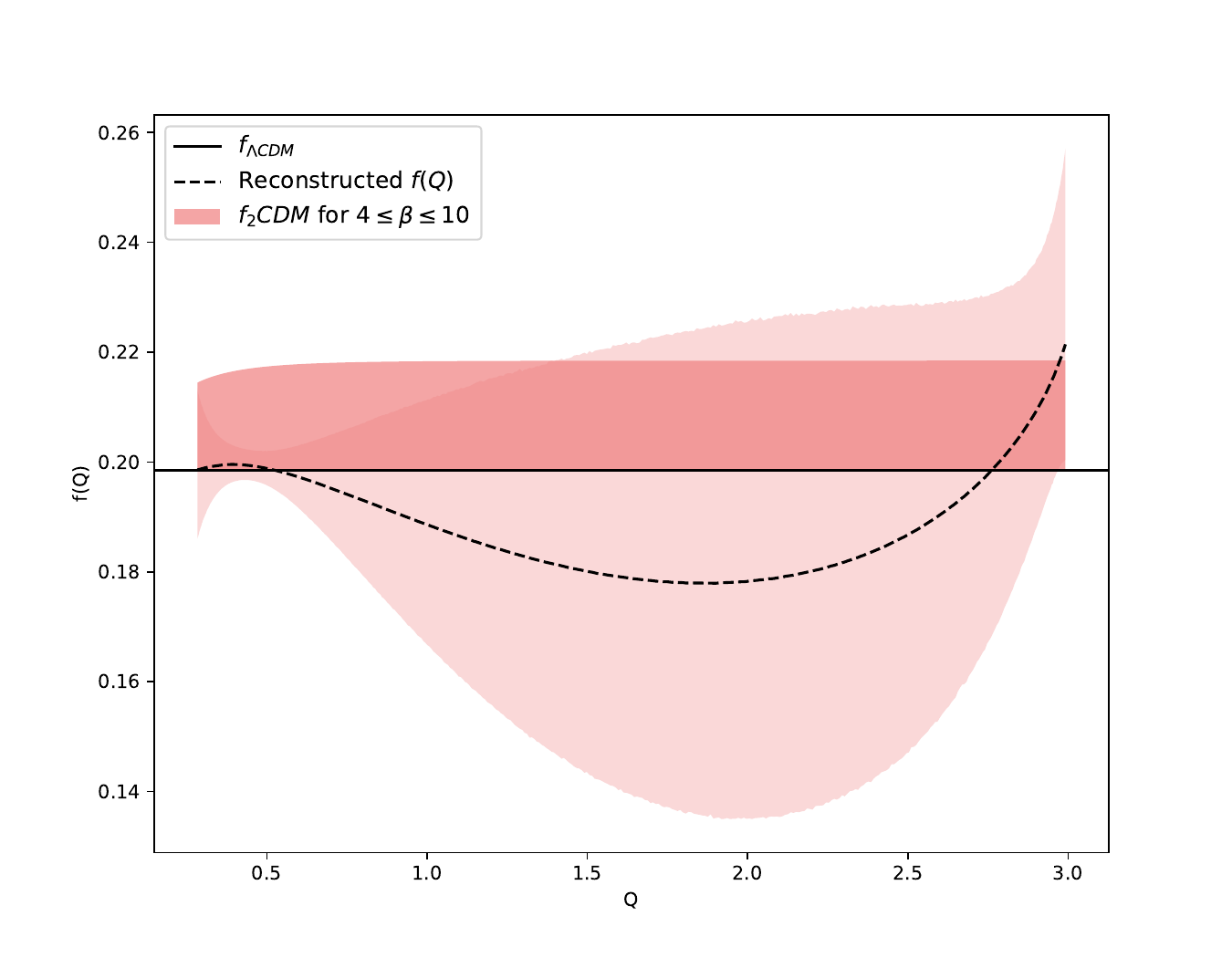}{0.5\textwidth}{(b)}
          }
\caption{The reconstructed behaviour for $f(Q)$ as a function of $Q$, resulting from data-driven reconstructions of $H(z)$ and $H'(z)$. The black dashed line in the graph represents the mean reconstructed curve, while the light pink colored regions indicate $1\sigma$ errors arising from the GP errors. Moreover, the black solid line marks the scenario for the cosmological constant $f_{\Lambda CDM}=6H_0^2(1 -\Omega_{m0})$. $f(Q)$ and $Q$ are both expressed in $H^2(z)$ units of $(km\,\,s^{-1}\,\,\,\,Mpc^{-1})^2$, and we display them normalized by $10^5$. In the left panel (a), the dark pink region displays the $f_1CDM$ model for $-0.03\le n \le 0.03$. In the right panel (b), the dark pink region displays the $f_2CDM$ model for $4\le \beta \le 10$. To present the plots in a simplified manner, we divided them by a factor of $10^5$.  \label{Recf}}
\end{figure*}

In Figure \ref{Recf}, we present the reconstructed function $f(Q)$ using the GP against $Q$. Now, our aim find the appropriate functional form of $f(Q)$ from our results, which will able to mimic the reconstructed $f(Q)$.

In the reconstruction profile, we presented the mean reconstruction curve alongside the $\Lambda$CDM model depicted by the straight line. This line maintains a constant value of $2\Lambda=-19267$, derived from the analysis. Observably, the reconstructed mean curve does not hold a constant value like $\Lambda$, but rather embodies the best-fit curve from the Gaussian analysis. It adopts a second-order polynomial form as $f(Q)= -2\Lambda+ \eta Q+\epsilon Q^2$, with the parameter values $\eta\simeq-1.45\times 10^{-3}$ and $\epsilon\simeq5.05\times 10^{-9}$, constrained by the reconstructed $f(Q)$ data. Notably, the functional form of $f(Q)$ simplifies to $f(Q)= -2\Lambda+ \eta Q+\epsilon Q^2$, Consequently, the reconstructed functional form now relies solely on one parameter, $\epsilon$,  as the linear term merges with the standard linear form in the action. Although one could introduce additional parameters into the reconstructed function, but a model with fewer parameters typically represents a better model than one with more. Therefore, we adhere to the one free parameter form of $f(Q)$ denoted as
\begin{equation}
\label{24}
f(Q)= -2\Lambda+\epsilon\,Q^2,
\end{equation}
where $\epsilon$ is the sole free parameter with units of $km\,s^{-1}\,\,Mpc^{-1}$. Note that, if we used a dimensionless parameter, then we may rewrite the Eq.\eqref{24} as $f(Q)= -2\Lambda+\epsilon'\,\frac{Q^2}{Q_0^2}$, where the dimensionless parameter $\epsilon'$ is defined as $\epsilon'=36\,H_0^4\,\epsilon$.\\
Subsequently, the reconstructed curve $f(Q)$, along with its shaded error regions, aids in discerning the true form of some widely studied functions of $f(Q)$. To this end, we compare two $f(Q)$ models: a power law-type and an exponential type, in search of suitable functions.

\subsection{$f_1$CDM :- $f(Q)=\alpha \left(\frac{Q}{Q_0}\right)^{n}$}
First, we consider the power-law $f(Q)$ model ($f_1$CDM) \citep{Jim_nez_2020,Lazkoz_2019,Khyllep_2023}, which is of the form $f(Q)=\alpha \left(\frac{Q}{Q_0}\right)^{n}$, with $\alpha=\frac{(\Omega_{m0}-1)\,6H_0^2}{2n-1}$. 
When $n=0$, the model reduces to $f_{\Lambda CDM}=-2\Lambda=6H_0^2(1-\Omega_{m0})$, which recover the $\Lambda$CDM expansion history of the universe.\\
It's worth noting that any curve falling within the shaded area in Figure \ref{Recf} can be considered the true form of $f(Q)$, besides the mean reconstruction curve. Hence, we constrain the free parameter $n$ to determine which values of $n$ allow the $f_1$CDM model to fit within the reconstructed area. As shown in Figure \ref{Recf}a, the constraint value indicates that $n$ might fall between the range of $-0.03\le n \le 0.03$.\\
The DE equation of state parameter corresponding to $f_1$CDM is
\begin{equation}
    w_{DE}(z)=-1+\frac{2n}{3}\frac{(1+z)}{H(z)}\frac{dH(z)}{dz},
\end{equation}
and the deceleration parameter is 
\begin{equation}
    q(z)=-1+\frac{3}{2}\left[ \frac{H^2(z)+(\Omega_{m0}-1)H_0^2\left(\frac{H(z)}{H_0}\right)^{2n}}{H^2(z)+n\,(\Omega_{m0}-1)H_0^2\left(\frac{H(z)}{H_0}\right)^{2n}}\right].
\end{equation}
In Figures 3(a), 3(b), and 3(c), we present the reconstructed profiles of $\Omega_{DE}$, $\omega_{DE}$, and $q(z)$ for $f_{1}$CDM, respectively. The constraint values are shown in Table \ref{table 2}.
                    
\begin{figure*}
\gridline{\fig{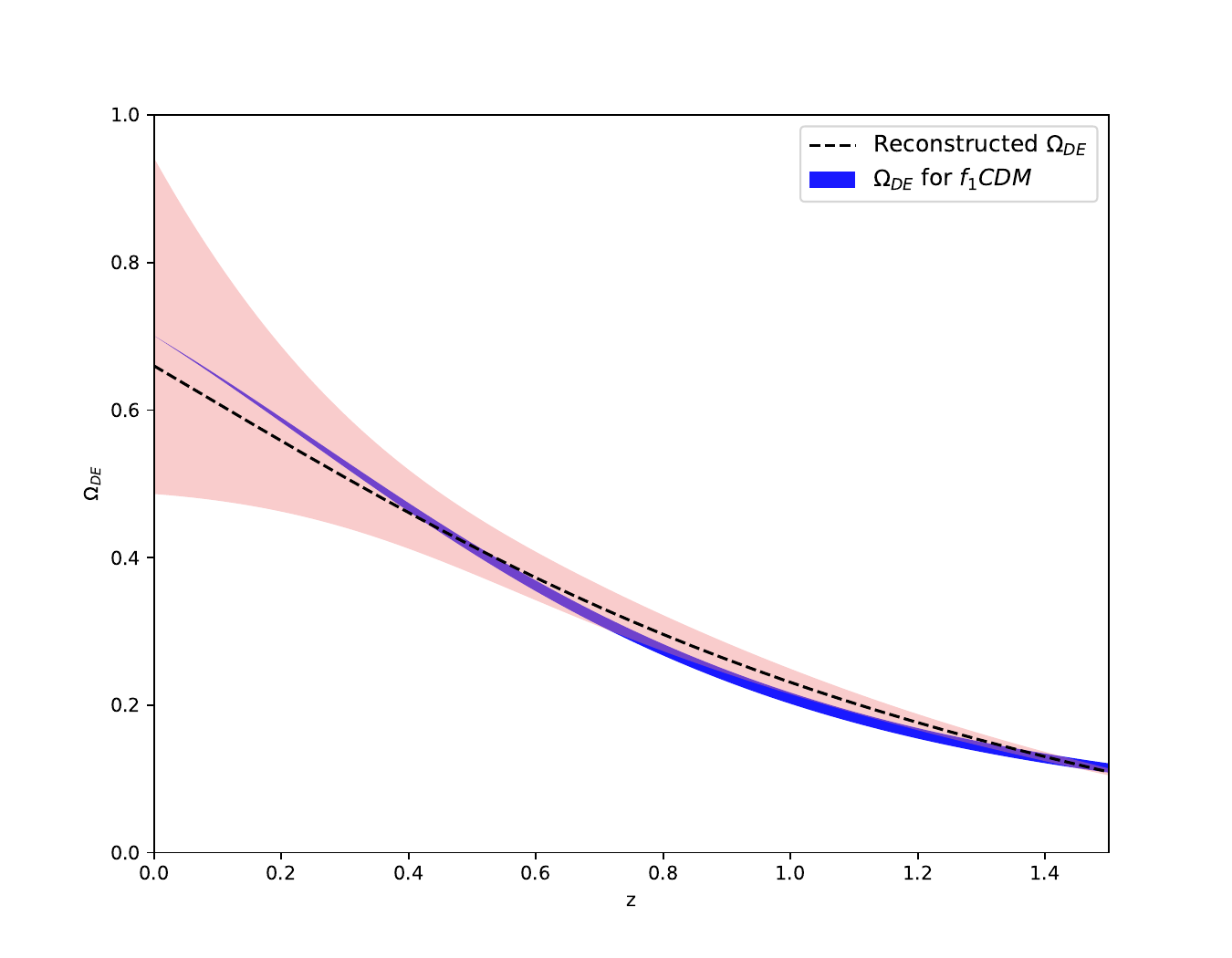}{0.48\textwidth}{(a)}
          \fig{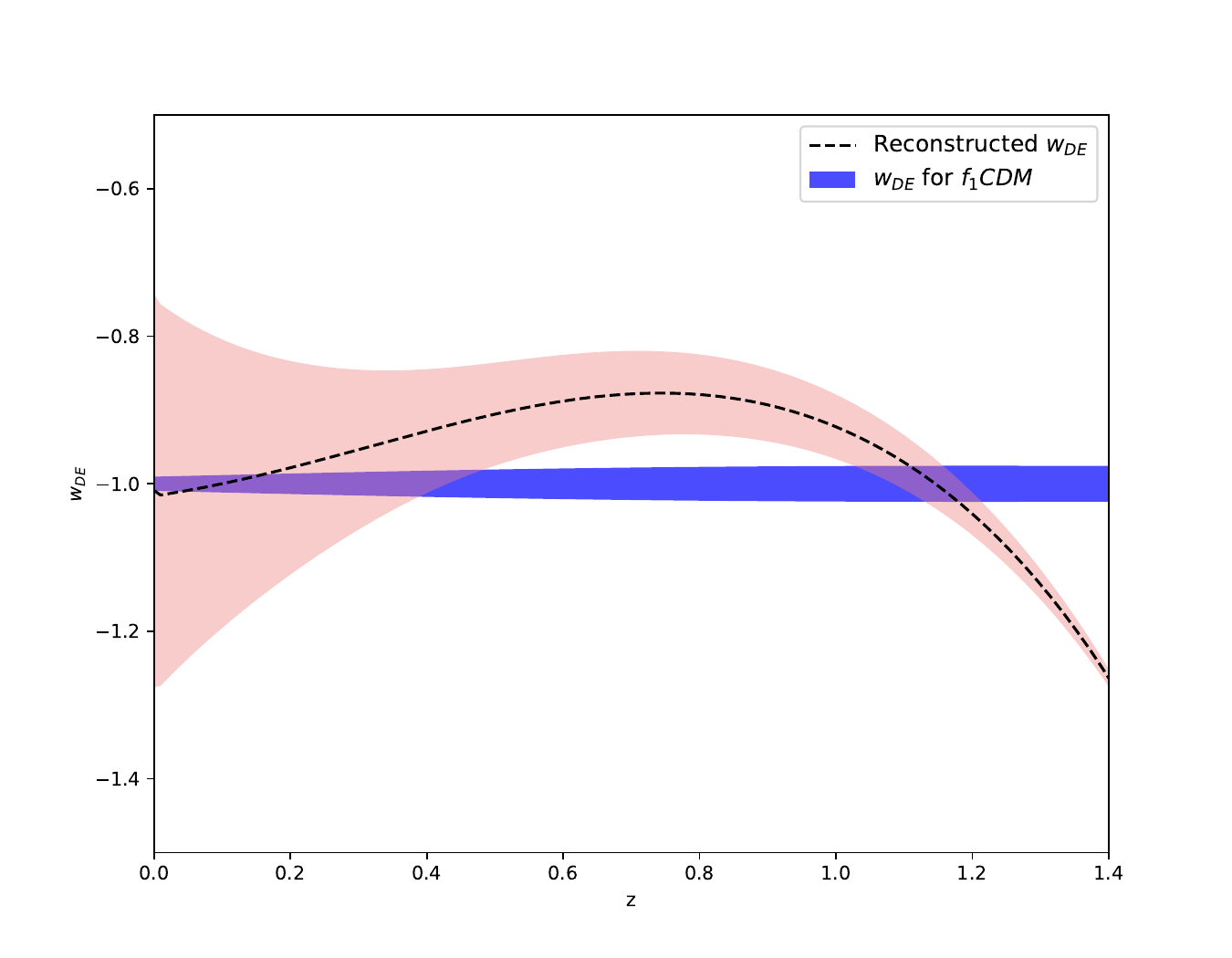}{0.48\textwidth}{(b)}
          }
\gridline{\fig{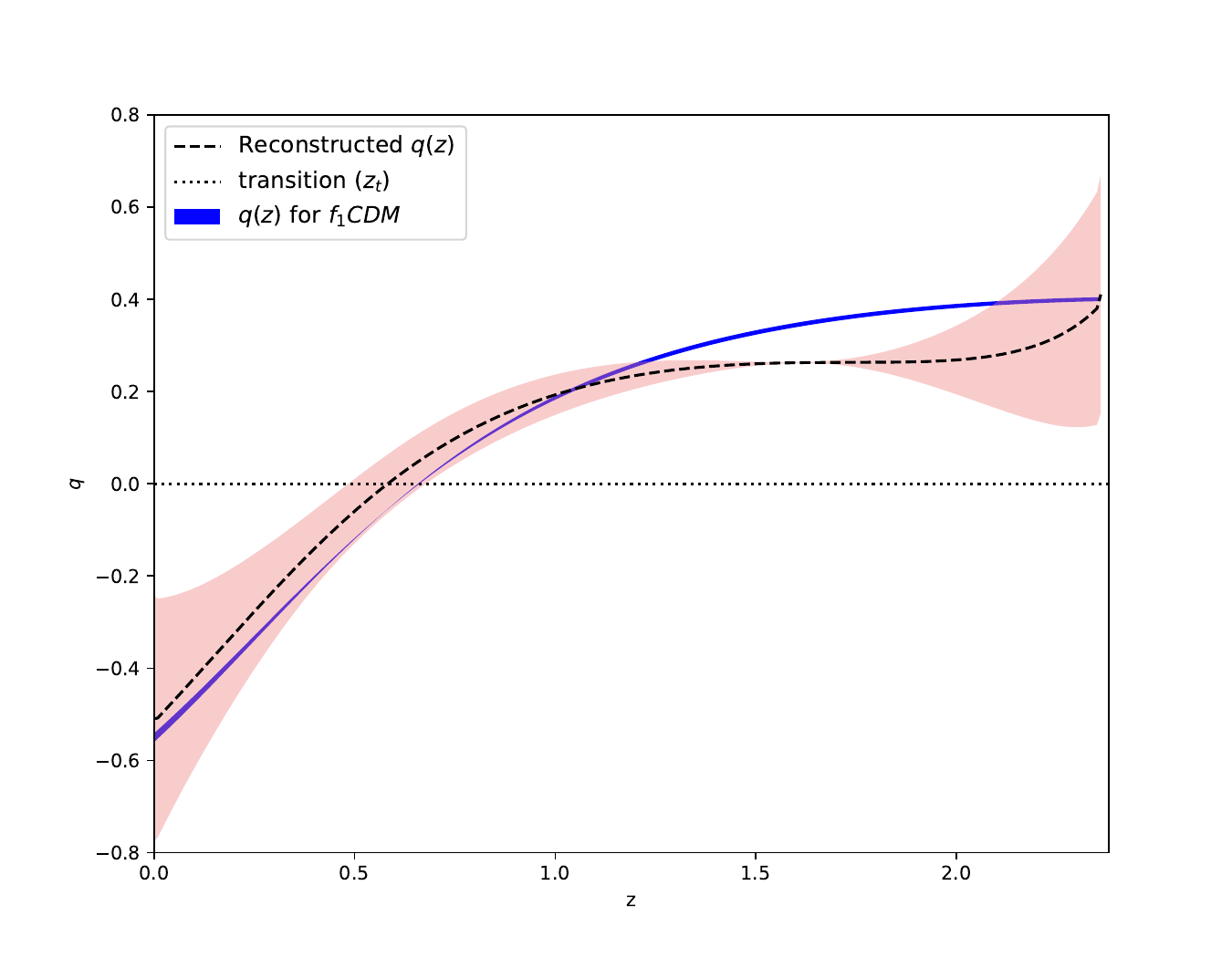}{0.48\textwidth}{(c)}}
\caption{The reconstructed forms of the dark energy density parameter $\Omega_{DE}$ (panel a), the dark energy equation-of-state parameter $w_{DE}$ (panel b), and the deceleration parameter $q$ (panel c) are derived using the reconstructed functions of $H(z)$, $H'(z)$, and the $f(Q)$, with obtained $H_0=68.71\pm 4.3$ $km\,s^{-1}\,\,Mpc^{-1}$ from the GP. In each graph, the black dashed line represents the mean reconstructed curve, while the shaded regions in different colors indicate the $1\sigma$ error resulting from the GP uncertainties. Furthermore, we have incorporated the projection of a plausible $f_1$CDM model by using reconstructed $H(z)$ and $H'(z)$ from GP and considering the range of the free parameter $-0.03 \leq n \leq 0.03$, depicted by a dark blue-shaded area in each graph.
\label{f1}}
\end{figure*}



\subsection{$f_2$CDM :- $f(Q)=\alpha Q_0\left(1-e^{-\beta\sqrt{\frac{Q}{Q_0}}}\right)$ }

Next, we consider the exponential $f(Q)$ model ($f_2$CDM) \citep{Sokoliuk_2023,Anagnostopoulos_2023,Khyllep_2023}, which is of the form $f(Q)=\alpha Q_0\left(1-e^{-\beta\sqrt{\frac{Q}{Q_0}}}\right)$, with $\alpha=\frac{1-\Omega_{m0}}{1-(1+\beta)e^{-\beta}}$. For $\beta=0$ the model reduces to the symmetric teleparallel theory equivalent to GR without a cosmological constant. When $\beta \to +\infty$, the model reduces to $f_{\Lambda CDM}=-2\Lambda=6H_0^2(1-\Omega_{m0})$, which recover the $\Lambda$CDM expansion history of the universe.\\
Here also, we constrain the free parameter $\beta$ to determine which values of $\beta$ allow the $f_2$CDM model to fit within the reconstructed area. As shown in Figure \ref{Recf}b, the constraint value indicates that $\beta$ might fall between the range of $4\le \beta \le 10$.\\
The DE equation of state parameter corresponding to $f_2$CDM is
\begin{equation}
    w_{DE}(z)=-1+\frac{\beta^2(1+z)H(z)}{3H_0\left(-H_0-\beta\,H(z)+H_0\,e^{\beta\frac{H(z)}{H_0}}\right)}\frac{dH(z)}{dz},
\end{equation}
and the deceleration parameter is 
\begin{equation}
    q(z)=-1+\frac{3}{H^2(z)}\left[\frac{H^2(z)+\frac{1-\Omega_{m0}}{1-(1+\beta)e^{-\beta}}H_0^2\left(-1+\left(1+\beta \frac{H(z)}{H_0}\right)e^{-\beta\frac{H(z)}{H_0}}\right)}{2-\frac{1-\Omega_{m0}}{1-(1+\beta)e^{-\beta}}\beta^2\,e^{-\beta\frac{H(z)}{H_0}}}\right].
\end{equation}
In Figures 4(a), 4(b), and 4(c), we present the reconstructed profiles of $\Omega_{DE}$,  $\omega_{DE}$, and $q(z)$ for  $f_{2}$CDM, respectively. The constraint values are shown in Table \ref{table 2}.
\begin{figure*}
\gridline{\fig{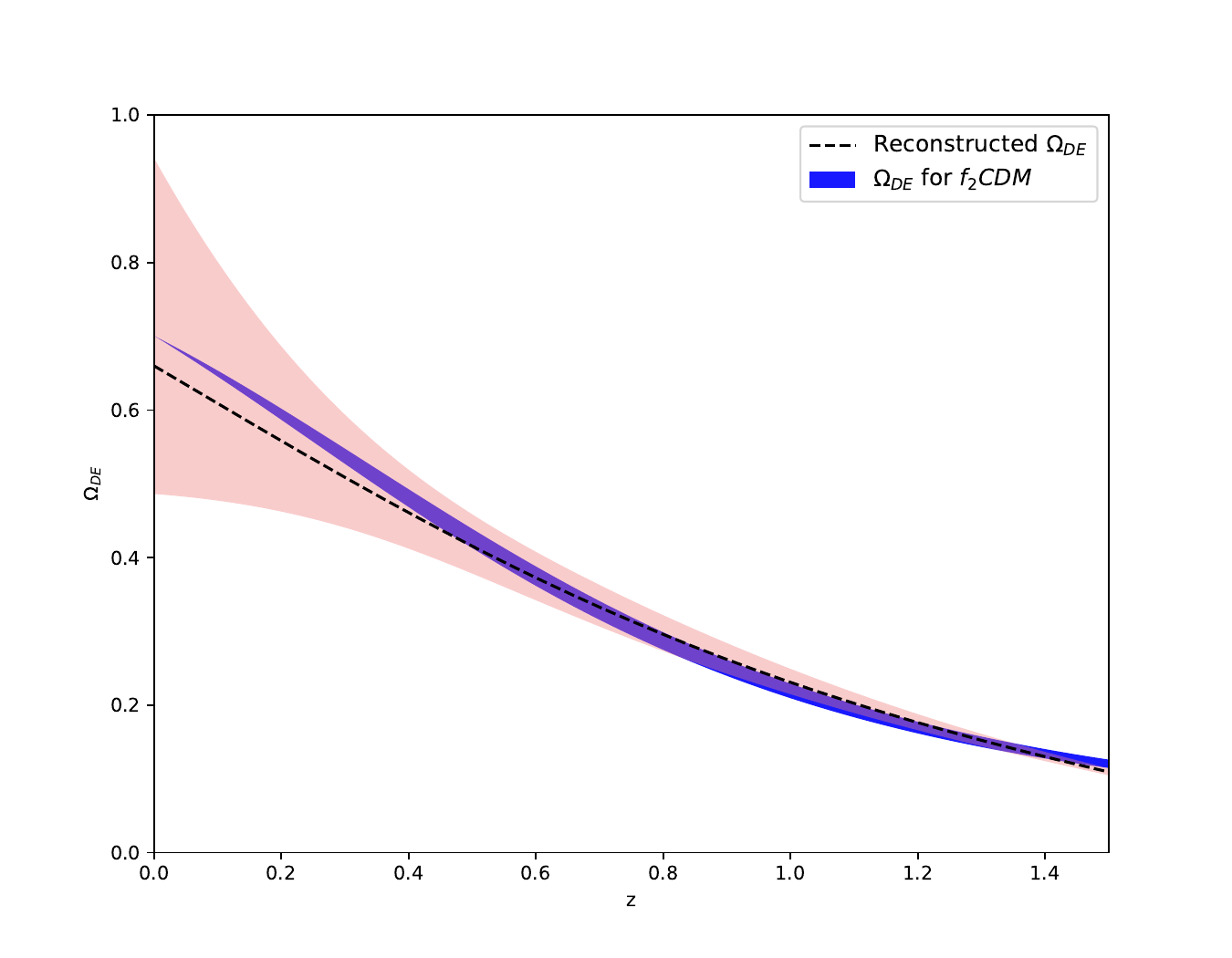}{0.48\textwidth}{(a)}
          \fig{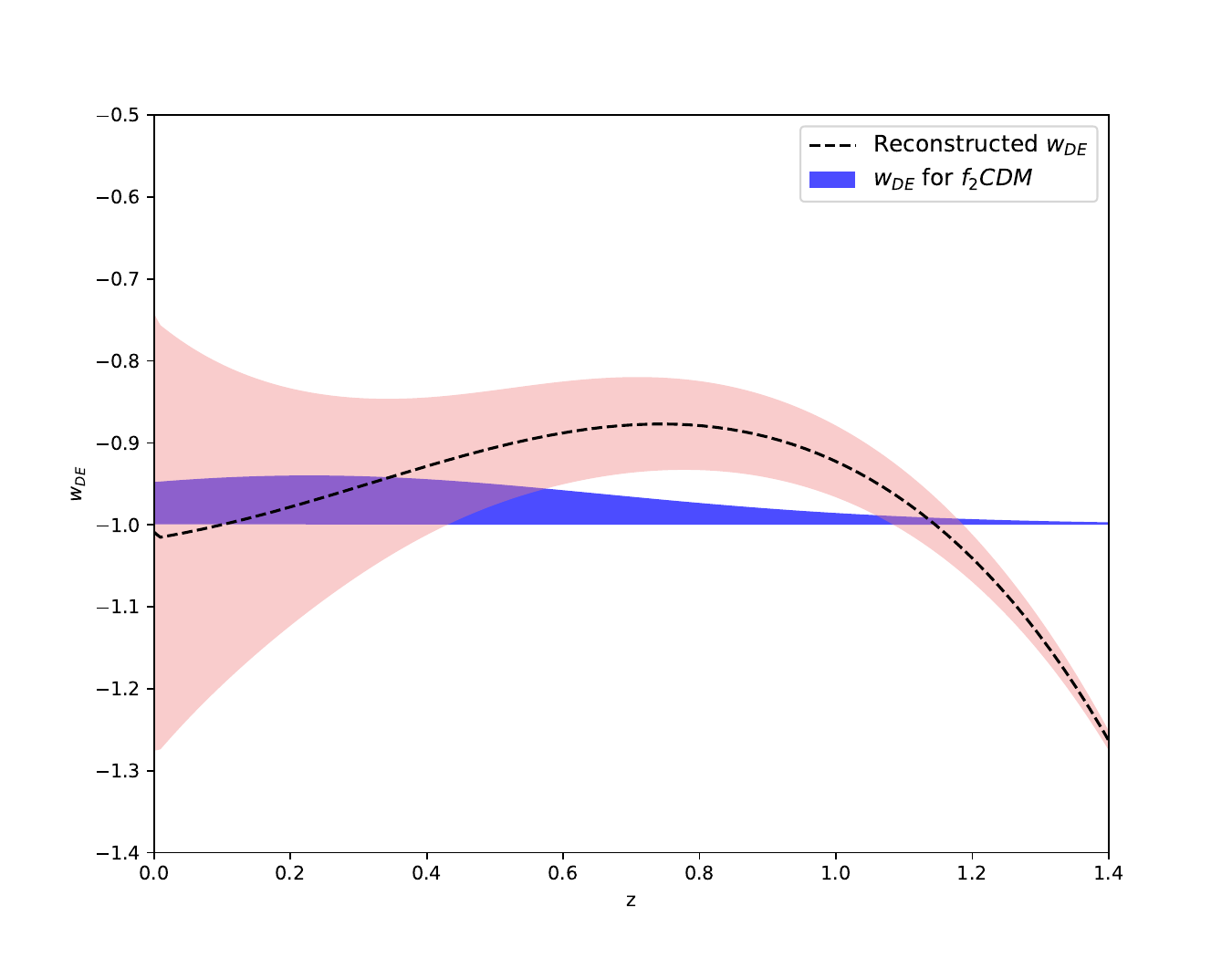}{0.48\textwidth}{(b)}
          }
\gridline{\fig{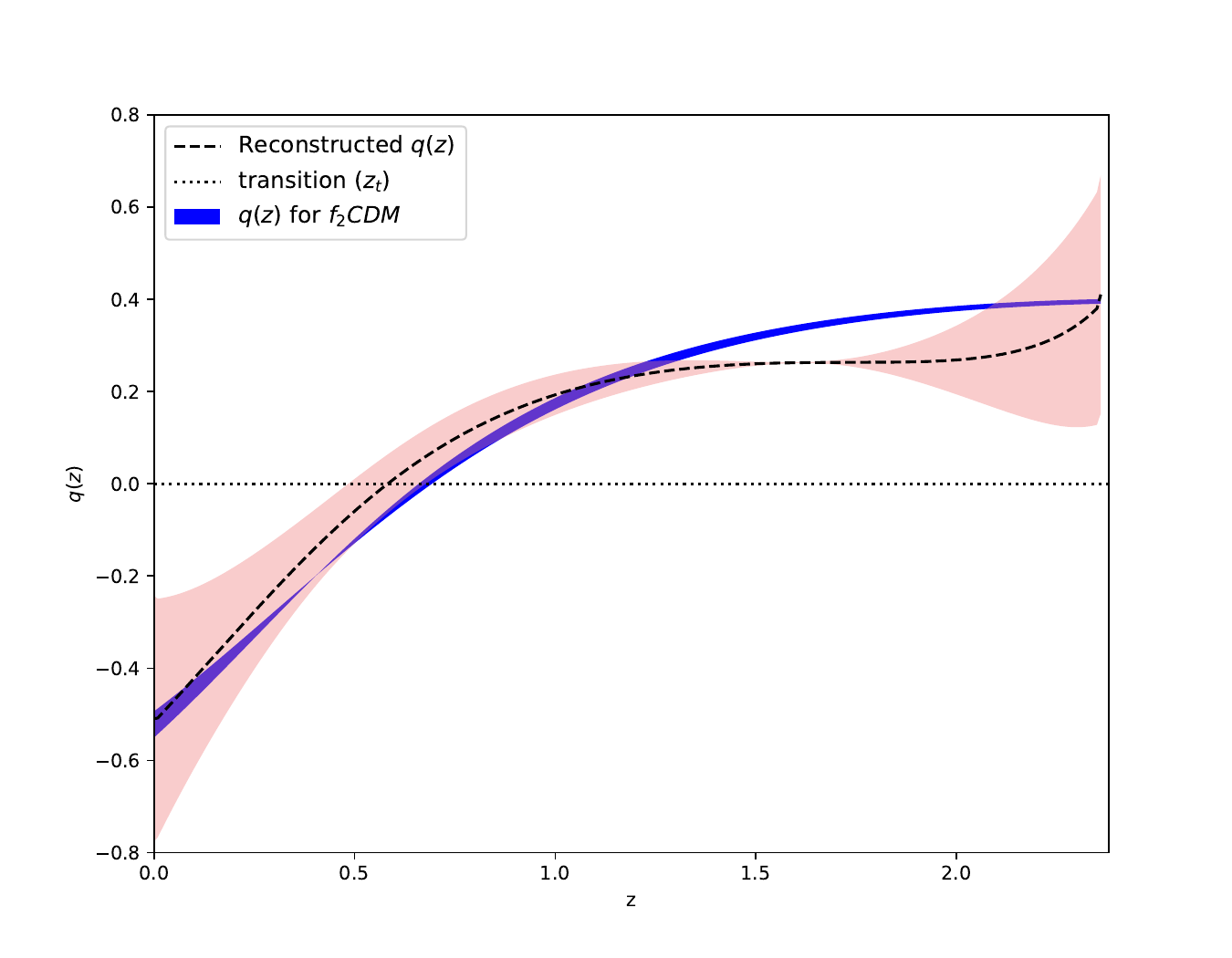}{0.48\textwidth}{(c)}}
\caption{Here, we have incorporated the projection of a plausible $f_2$CDM model by using reconstructed $H(z)$ and $H'(z)$ from GP and considering the range of the free parameter $4 \leq \beta \leq 10$, depicted by a dark blue-shaded area in each graph.
\label{f2}}
\end{figure*}

\begin{deluxetable*}{lccc}
\tablenum{2}
\tablecaption{Cosmographic parameters values have been acquired for the reconstructed $f(Q)$ model, as well as for the $f_1$CDM and $f_2$CDM models.\label{table 2}}
\tablewidth{0pt}
\tablehead{
\colhead{Model} & \colhead{$\Omega_{DE}$} & \colhead{$w_{DE}$} & \colhead{$q_{0}$}  
}
\decimalcolnumbers
\startdata
    Reconstructed $f(Q)$       &  $0.661^{+0.279}_{-0.176}$ & $-1.0089^{+0.2689}_{-0.2681}$ & $-0.5091^{+0.2641}_{-0.2659}$\\[1ex]
    $f_1$CDM   &  $0.7$ &  $-1.0113\le w_{DE}\le-0.9887$ & $-0.5593\le q_0\le-0.5403$  \\[1ex]
$f_2$CDM &  $0.7$ &  $-1\le w_{DE}\le -0.9462$ & $-0.55\le q_0\le -0.494$ \\[1ex]
\enddata
\end{deluxetable*}

\begin{figure*}
\gridline{\fig{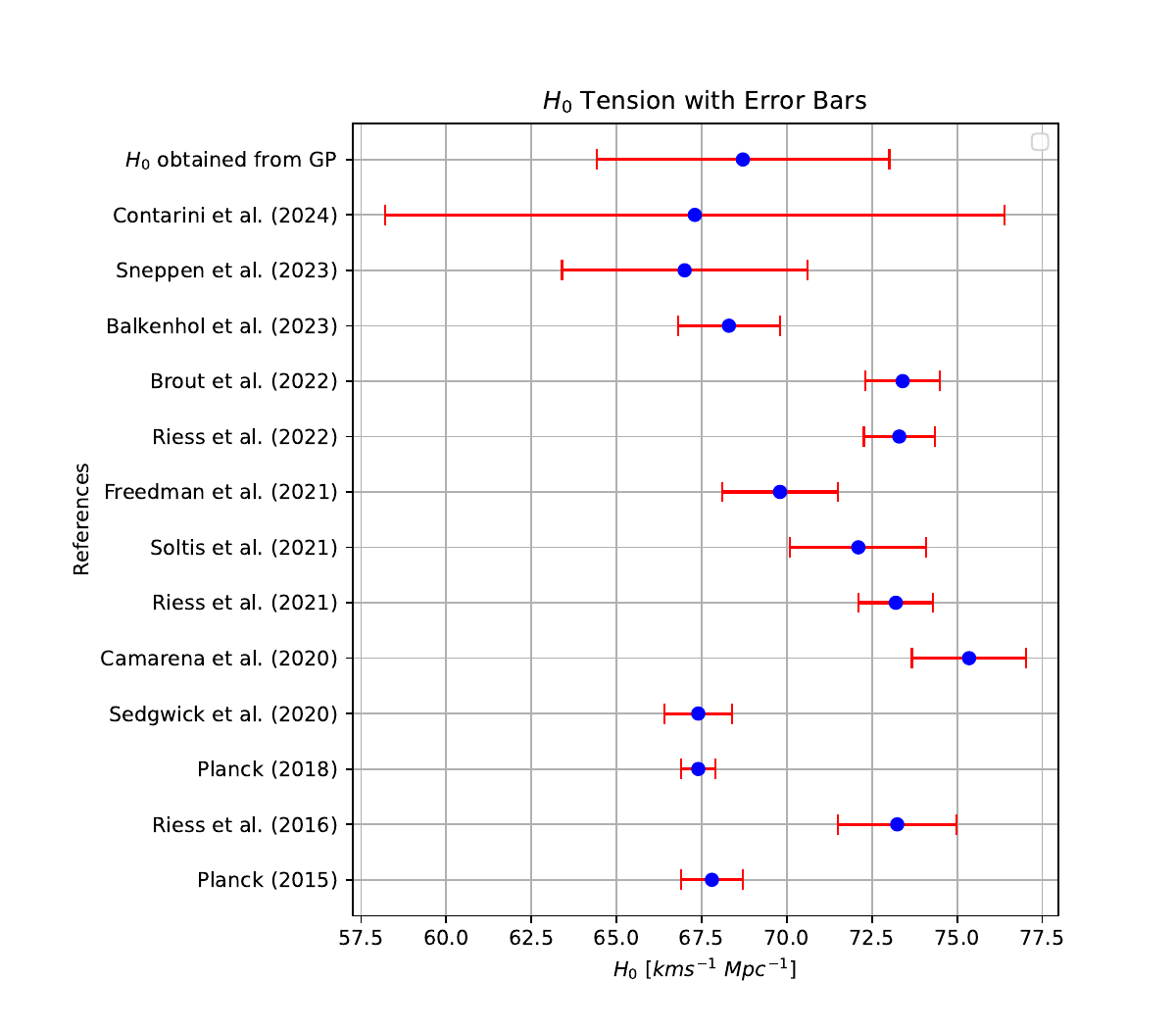}{0.5\textwidth}{}
           }
\caption{Recent estimations of the Hubble constant $H_0$. The reconstructed $H(z)$ from Gaussian processes using the OHD dataset gives the value $H_0=68.71\pm 4.3$ $km\,s^{-1}\,\,Mpc^{-1}$.
\label{H0}}
\end{figure*}

\section{Conclusion}
\label{section 5}
In this manuscript, we have independently reconstructed the \( f(Q) \) function using observational measurements. To achieve this objective, we have incorporated local Hubble measurements, including Cosmic Chronometers and Baryon Acoustic Oscillations (BAO), and employed Gaussian Processes (GP) for statistical analysis. Recent inquiries into modified gravities have spurred the search for a function derivable from observational data. Typically, researchers assume specific functional forms for \( f(Q) \) and then constrain the free parameters using observational measurements, often arbitrary assumptions. However, the GP methodology allows us to discern the functional form of \( f(Q) \) independently, without imposing specific conditions based on observational measurements.

Our analysis encompasses Hubble measurements, encompassing Cosmic Chronometers and BAO measurements, for GP analysis. From this scrutiny, we determined the value of \( H_0 \), which not only resolves the \( H_0 \) tension issue in a model-independent manner but also aligns closely with recent precise studies on the subject. To advance our investigation, we first reconstructed \( H(z) \) and its first derivative ${H'(z)}$ from observational samples. Given that the non-metric scalar \( Q \) is a function of \( H \), all Friedmann equations can be expressed in terms of \( H(z) \) and its first derivative ${H'(z)}$. Leveraging the reconstructed functions of \( H \) and its derivative $H'$, we reconstructed \( f(Q) \) without resorting to any assumptions. This reconstructed \( f(Q) \) addresses the \( H_0 \) issue by employing local Hubble measurements. We have presented the recent estimates of the Hubble constant $H_0$ along with its constraint value from our GP analysis in Figure \ref{H0}.

Figure \ref{Recf} displays the profile of the reconstructed function \( f(Q) \) concerning the non-metricity scalar \( Q \), where the dark dotted line represents the mean reconstructed function, the shaded region denotes the error, and the black line indicates the $\Lambda$CDM model. The deviation of the reconstructed function from $\Lambda$CDM suggests a quadratic behavior, leading us to propose a quadratic functional form of \( f(Q) \) with a single free parameter, quantifying the deviation from $\Lambda$CDM as $f(Q)= -2\Lambda+\epsilon Q^2$. Furthermore, we constrain the range of this free parameter within which the one-parameter \( f(Q) \) function lies in the reconstructed \( f(Q) \) region given by $-4.809 \times 10^{-9}< \epsilon < 5.658 \times 10^{-10}$.

Additionally, we scrutinized two widely studied forms of \( f(Q) \) against the reconstructed \(f(Q) \) function, enhancing the constraint on the free parameters compared to traditional observational constraints, and presented the improved parameter range. Moreover, we explore cosmological implications, investigating deceleration parameters, dimensionless dark energy, and the dark energy equation of state parameters for specific models. As anticipated, our findings corroborate the current accelerated expansion of the universe, consistent with recent studies.

In conclusion, our study presents a model-independent reconstruction and proposition of the \( f(Q) \) functional form, solely relying on observational measurements through GP analysis. This approach not only enhances constraints on the free parameters of specific models but also circumvents arbitrary choices for gravitational Lagrangian functions. While our study focuses on Hubble measurements, future endeavors could extend this analysis to other observational measurements, such as supernovae, which we aspire to explore in forthcoming research.\\

\textbf{Data availability:} There are no new data associated with this article.\\

\begin{acknowledgments}  GNG acknowledges the University Grants Commission (UGC), New Delhi, India, for awarding a Senior Research Fellowship (UGC-Ref. No.: 201610122060). SM acknowledges the Japan Society for the Promotion of Science (JSPS) for providing postdoctoral felowship during 2024-2026 (JSPS ID No.: P24026). PKS acknowledges Science and Engineering Research Board, Department of Science and Technology, Government of India for financial support to carry out Research project No.: CRG/2022/001847 and IUCAA, Pune, India for providing support through the visiting Associateship program. We are very much grateful to the honorable referee and to the editor for the illuminating suggestions that have significantly improved our work in terms of research quality, and presentation.
\end{acknowledgments}


@article{Capozziello_2011,
   title={Extended Theories of Gravity},
   volume={509},
   ISSN={0370-1573},
   url={http://dx.doi.org/10.1016/j.physrep.2011.09.003},
   DOI={10.1016/j.physrep.2011.09.003},
   number={4–5},
   journal={Physics Reports},
   publisher={Elsevier BV},
   author={Capozziello, Salvatore and De Laurentis, Mariafelicia},
   year={2011},
   month=dec, pages={167–321} }

@article{Nojiri_2011,
   title={Unified cosmic history in modified gravity: From <mml:math xmlns:mml="http://www.w3.org/1998/Math/MathML" altimg="si10.gif" display="inline" overflow="scroll"><mml:mi>F</mml:mi><mml:mrow><mml:mo>(</mml:mo><mml:mi>R</mml:mi><mml:mo>)</mml:mo></mml:mrow></mml:math> theory to Lorentz non-invariant models},
   volume={505},
   ISSN={0370-1573},
   url={http://dx.doi.org/10.1016/j.physrep.2011.04.001},
   DOI={10.1016/j.physrep.2011.04.001},
   number={2–4},
   journal={Physics Reports},
   publisher={Elsevier BV},
   author={Nojiri, Shin’ichi and Odintsov, Sergei D.},
   year={2011},
   month=aug, pages={59–144} }

@article{Peebles_2003,
   title={The cosmological constant and dark energy},
   volume={75},
   ISSN={1539-0756},
   url={http://dx.doi.org/10.1103/RevModPhys.75.559},
   DOI={10.1103/revmodphys.75.559},
   number={2},
   journal={Reviews of Modern Physics},
   publisher={American Physical Society (APS)},
   author={Peebles, P. J. E. and Ratra, Bharat},
   year={2003},
   month=apr, pages={559–606} }

@article{Cai_2010,
   title={Quintom cosmology: Theoretical implications and observations},
   volume={493},
   ISSN={0370-1573},
   url={http://dx.doi.org/10.1016/j.physrep.2010.04.001},
   DOI={10.1016/j.physrep.2010.04.001},
   number={1},
   journal={Physics Reports},
   publisher={Elsevier BV},
   author={Cai, Yi-Fu and Saridakis, Emmanuel N. and Setare, Mohammad R. and Xia, Jun-Qing},
   year={2010},
   month=aug, pages={1–60} }

@article{Li_2019,
   title={A Simple Phenomenological Emergent Dark Energy Model can Resolve the Hubble Tension},
   volume={883},
   ISSN={2041-8213},
   url={http://dx.doi.org/10.3847/2041-8213/ab3e09},
   DOI={10.3847/2041-8213/ab3e09},
   number={1},
   journal={The Astrophysical Journal Letters},
   publisher={American Astronomical Society},
   author={Li, Xiaolei and Shafieloo, Arman},
   year={2019},
   month=sep, pages={L3} }

@article{Li_2019_2,
   title={Revisiting Metastable Dark Energy and Tensions in the Estimation of Cosmological Parameters},
   volume={887},
   ISSN={1538-4357},
   url={http://dx.doi.org/10.3847/1538-4357/ab535d},
   DOI={10.3847/1538-4357/ab535d},
   number={2},
   journal={The Astrophysical Journal},
   publisher={American Astronomical Society},
   author={Li, Xiaolei and Shafieloo, Arman and Sahni, Varun and Starobinsky, Alexei A.},
   year={2019},
   month=dec, pages={153} }

@article{Elizalde_2019,
   title={Cosmological singularities in interacting dark energy models with an ω(q) parametrization},
   volume={28},
   ISSN={1793-6594},
   url={http://dx.doi.org/10.1142/S0218271819500196},
   DOI={10.1142/s0218271819500196},
   number={01},
   journal={International Journal of Modern Physics D},
   publisher={World Scientific Pub Co Pte Lt},
   author={Elizalde, Emilio and Khurshudyan, Martiros and Nojiri, Shin’ichi},
   year={2019},
   month=jan, pages={1950019} }

@article{Wong_2019,
   title={H0LiCOW – XIII. A 2.4 per cent measurement of H0 from lensed quasars: 5.3σ tension between early- and late-Universe probes},
   volume={498},
   ISSN={1365-2966},
   url={http://dx.doi.org/10.1093/mnras/stz3094},
   DOI={10.1093/mnras/stz3094},
   number={1},
   journal={Monthly Notices of the Royal Astronomical Society},
   publisher={Oxford University Press (OUP)},
   author={Wong, Kenneth C and Suyu, Sherry H and Chen, Geoff C-F and Rusu, Cristian E and Millon, Martin and Sluse, Dominique and Bonvin, Vivien and Fassnacht, Christopher D and Taubenberger, Stefan and Auger, Matthew W and Birrer, Simon and Chan, James H H and Courbin, Frederic and Hilbert, Stefan and Tihhonova, Olga and Treu, Tommaso and Agnello, Adriano and Ding, Xuheng and Jee, Inh and Komatsu, Eiichiro and Shajib, Anowar J and Sonnenfeld, Alessandro and Blandford, Roger D and Koopmans, Léon V E and Marshall, Philip J and Meylan, Georges},
   year={2019},
   month=sep, pages={1420–1439} }

@article{Freedman_2019,
   title={The Carnegie-Chicago Hubble Program. VIII. An Independent Determination of the Hubble Constant Based on the Tip of the Red Giant Branch*},
   volume={882},
   ISSN={1538-4357},
   url={http://dx.doi.org/10.3847/1538-4357/ab2f73},
   DOI={10.3847/1538-4357/ab2f73},
   number={1},
   journal={The Astrophysical Journal},
   publisher={American Astronomical Society},
   author={Freedman, Wendy L. and Madore, Barry F. and Hatt, Dylan and Hoyt, Taylor J. and Jang, In Sung and Beaton, Rachael L. and Burns, Christopher R. and Lee, Myung Gyoon and Monson, Andrew J. and Neeley, Jillian R. and Phillips, M. M. and Rich, Jeffrey A. and Seibert, Mark},
   year={2019},
   month=aug, pages={34} }

@article{Vagnozzi_2020,
   title={New physics in light of the 
<mml:math xmlns:mml="http://www.w3.org/1998/Math/MathML" display="inline"><mml:msub><mml:mi>H</mml:mi><mml:mn>0</mml:mn></mml:msub></mml:math>
 tension: An alternative view},
   volume={102},
   ISSN={2470-0029},
   url={http://dx.doi.org/10.1103/PhysRevD.102.023518},
   DOI={10.1103/physrevd.102.023518},
   number={2},
   journal={Physical Review D},
   publisher={American Physical Society (APS)},
   author={Vagnozzi, Sunny},
   year={2020},
   month=jul }

@article{Cai_2016,
   title={f(T) teleparallel gravity and cosmology},
   volume={79},
   ISSN={1361-6633},
   url={http://dx.doi.org/10.1088/0034-4885/79/10/106901},
   DOI={10.1088/0034-4885/79/10/106901},
   number={10},
   journal={Reports on Progress in Physics},
   publisher={IOP Publishing},
   author={Cai, Yi-Fu and Capozziello, Salvatore and De Laurentis, Mariafelicia and Saridakis, Emmanuel N},
   year={2016},
   month=sep, pages={106901} }

@article{Jim_nez_2018,
   title={Coincident general relativity},
   volume={98},
   ISSN={2470-0029},
   url={http://dx.doi.org/10.1103/PhysRevD.98.044048},
   DOI={10.1103/physrevd.98.044048},
   number={4},
   journal={Physical Review D},
   publisher={American Physical Society (APS)},
   author={Jiménez, Jose Beltrán and Heisenberg, Lavinia and Koivisto, Tomi},
   year={2018},
   month=aug }

@article{Lazkoz_2019,
   title={Observational constraints of 
<mml:math xmlns:mml="http://www.w3.org/1998/Math/MathML" display="inline"><mml:mi>f</mml:mi><mml:mo mathvariant="bold" stretchy="false">(</mml:mo><mml:mi>Q</mml:mi><mml:mo mathvariant="bold" stretchy="false">)</mml:mo></mml:math>
 gravity},
   volume={100},
   ISSN={2470-0029},
   url={http://dx.doi.org/10.1103/PhysRevD.100.104027},
   DOI={10.1103/physrevd.100.104027},
   number={10},
   journal={Physical Review D},
   publisher={American Physical Society (APS)},
   author={Lazkoz, Ruth and Lobo, Francisco S. N. and Ortiz-Baños, María and Salzano, Vincenzo},
   year={2019},
   month=nov }

@article{Anagnostopoulos_2021,
   title={First evidence that non-metricity f(Q) gravity could challenge ΛCDM},
   volume={822},
   ISSN={0370-2693},
   url={http://dx.doi.org/10.1016/j.physletb.2021.136634},
   DOI={10.1016/j.physletb.2021.136634},
   journal={Physics Letters B},
   publisher={Elsevier BV},
   author={Anagnostopoulos, Fotios K. and Basilakos, Spyros and Saridakis, Emmanuel N.},
   year={2021},
   month=nov, pages={136634} }

@article{Frusciante_2021,
   title={Signatures of 
<mml:math xmlns:mml="http://www.w3.org/1998/Math/MathML" display="inline"><mml:mi>f</mml:mi><mml:mo stretchy="false">(</mml:mo><mml:mi>Q</mml:mi><mml:mo stretchy="false">)</mml:mo></mml:math>
 gravity in cosmology},
   volume={103},
   ISSN={2470-0029},
   url={http://dx.doi.org/10.1103/PhysRevD.103.044021},
   DOI={10.1103/physrevd.103.044021},
   number={4},
   journal={Physical Review D},
   publisher={American Physical Society (APS)},
   author={Frusciante, Noemi},
   year={2021},
   month=feb }

@article{Lin_2021,
   title={Spherically symmetric configuration in 
<mml:math xmlns:mml="http://www.w3.org/1998/Math/MathML" display="inline"><mml:mi>f</mml:mi><mml:mo stretchy="false">(</mml:mo><mml:mi>Q</mml:mi><mml:mo stretchy="false">)</mml:mo></mml:math>
 gravity},
   volume={103},
   ISSN={2470-0029},
   url={http://dx.doi.org/10.1103/PhysRevD.103.124001},
   DOI={10.1103/physrevd.103.124001},
   number={12},
   journal={Physical Review D},
   publisher={American Physical Society (APS)},
   author={Lin, Rui-Hui and Zhai, Xiang-Hua},
   year={2021},
   month=jun }

@article{Mandal_2020,
   title={Energy conditions in 
<mml:math xmlns:mml="http://www.w3.org/1998/Math/MathML" display="inline"><mml:mi>f</mml:mi><mml:mo stretchy="false">(</mml:mo><mml:mi>Q</mml:mi><mml:mo stretchy="false">)</mml:mo></mml:math>
 gravity},
   volume={102},
   ISSN={2470-0029},
   url={http://dx.doi.org/10.1103/PhysRevD.102.024057},
   DOI={10.1103/physrevd.102.024057},
   number={2},
   journal={Physical Review D},
   publisher={American Physical Society (APS)},
   author={Mandal, Sanjay and Sahoo, P. K. and Santos, J. R. L.},
   year={2020},
   month=jul }

@article{Mandal_2020_2,
   title={Cosmography in 
<mml:math xmlns:mml="http://www.w3.org/1998/Math/MathML" display="inline"><mml:mi>f</mml:mi><mml:mo stretchy="false">(</mml:mo><mml:mi>Q</mml:mi><mml:mo stretchy="false">)</mml:mo></mml:math>
 gravity},
   volume={102},
   ISSN={2470-0029},
   url={http://dx.doi.org/10.1103/PhysRevD.102.124029},
   DOI={10.1103/physrevd.102.124029},
   number={12},
   journal={Physical Review D},
   publisher={American Physical Society (APS)},
   author={Mandal, Sanjay and Wang, Deng and Sahoo, P. K.},
   year={2020},
   month=dec }

@article{Jim_nez_2020,
   title={Cosmology in 
<mml:math xmlns:mml="http://www.w3.org/1998/Math/MathML" display="inline"><mml:mi>f</mml:mi><mml:mo stretchy="false">(</mml:mo><mml:mi>Q</mml:mi><mml:mo stretchy="false">)</mml:mo></mml:math>
 geometry},
   volume={101},
   ISSN={2470-0029},
   url={http://dx.doi.org/10.1103/PhysRevD.101.103507},
   DOI={10.1103/physrevd.101.103507},
   number={10},
   journal={Physical Review D},
   publisher={American Physical Society (APS)},
   author={Jiménez, Jose Beltrán and Heisenberg, Lavinia and Koivisto, Tomi and Pekar, Simon},
   year={2020},
   month=may }

@article{Harko_2018,
   title={Coupling matter in modified<mml:math xmlns:mml="http://www.w3.org/1998/Math/MathML" display="inline"><mml:mi>Q</mml:mi></mml:math>gravity},
   volume={98},
   ISSN={2470-0029},
   url={http://dx.doi.org/10.1103/PhysRevD.98.084043},
   DOI={10.1103/physrevd.98.084043},
   number={8},
   journal={Physical Review D},
   publisher={American Physical Society (APS)},
   author={Harko, Tiberiu and Koivisto, Tomi S. and Lobo, Francisco S. N. and Olmo, Gonzalo J. and Rubiera-Garcia, Diego},
   year={2018},
   month=oct }

@misc{heisenberg2023review,
      title={Review on $f(Q)$ Gravity}, 
      author={Lavinia Heisenberg},
      year={2023},
      volume={79},
      url={https://doi.org/10.1016/j.physrep.2024.02.001},
      journal={Physics Reports},
}

@article{Xu_2019,
   title={f(Q,T) gravity},
   volume={79},
   ISSN={1434-6052},
   url={http://dx.doi.org/10.1140/epjc/s10052-019-7207-4},
   DOI={10.1140/epjc/s10052-019-7207-4},
   number={8},
   journal={The European Physical Journal C},
   publisher={Springer Science and Business Media LLC},
   author={Xu, Yixin and Li, Guangjie and Harko, Tiberiu and Liang, Shi-Dong},
   year={2019},
   month=aug }

@article{Capozziello_2023,
   title={The role of the boundary term in f(Q, B) symmetric teleparallel gravity},
   volume={83},
   ISSN={1434-6052},
   url={http://dx.doi.org/10.1140/epjc/s10052-023-12072-y},
   DOI={10.1140/epjc/s10052-023-12072-y},
   number={10},
   journal={The European Physical Journal C},
   publisher={Springer Science and Business Media LLC},
   author={Capozziello, Salvatore and De Falco, Vittorio and Ferrara, Carmen},
   year={2023},
   month=oct }


@article{De_2024,
   title={Non-metricity with boundary terms: ��(��,��) gravity and cosmology},
   volume={2024},
   ISSN={1475-7516},
   url={http://dx.doi.org/10.1088/1475-7516/2024/03/050},
   DOI={10.1088/1475-7516/2024/03/050},
   number={03},
   journal={Journal of Cosmology and Astroparticle Physics},
   publisher={IOP Publishing},
   author={De, Avik and Loo, Tee-How and Saridakis, Emmanuel N.},
   year={2024},
   month=mar, pages={050} }

@article{Ayuso_2021,
   title={Observational constraints on cosmological solutions of 
<mml:math xmlns:mml="http://www.w3.org/1998/Math/MathML" display="inline"><mml:mi>f</mml:mi><mml:mo stretchy="false">(</mml:mo><mml:mi>Q</mml:mi><mml:mo stretchy="false">)</mml:mo></mml:math>
 theories},
   volume={103},
   ISSN={2470-0029},
   url={http://dx.doi.org/10.1103/PhysRevD.103.063505},
   DOI={10.1103/physrevd.103.063505},
   number={6},
   journal={Physical Review D},
   publisher={American Physical Society (APS)},
   author={Ayuso, Ismael and Lazkoz, Ruth and Salzano, Vincenzo},
   year={2021},
   month=mar }

@article{Khyllep_2021,
   title={Cosmological solutions and growth index of matter perturbations in 
<mml:math xmlns:mml="http://www.w3.org/1998/Math/MathML" display="inline"><mml:mi>f</mml:mi><mml:mo stretchy="false">(</mml:mo><mml:mi>Q</mml:mi><mml:mo stretchy="false">)</mml:mo></mml:math>
 gravity},
   volume={103},
   ISSN={2470-0029},
   url={http://dx.doi.org/10.1103/PhysRevD.103.103521},
   DOI={10.1103/physrevd.103.103521},
   number={10},
   journal={Physical Review D},
   publisher={American Physical Society (APS)},
   author={Khyllep, Wompherdeiki and Paliathanasis, Andronikos and Dutta, Jibitesh},
   year={2021},
   month=may }

@article{Khyllep_2023,
   title={Cosmology in 
<mml:math xmlns:mml="http://www.w3.org/1998/Math/MathML" display="inline"><mml:mi>f</mml:mi><mml:mo stretchy="false">(</mml:mo><mml:mi>Q</mml:mi><mml:mo stretchy="false">)</mml:mo></mml:math>
 gravity: A unified dynamical systems analysis of the background and perturbations},
   volume={107},
   ISSN={2470-0029},
   url={http://dx.doi.org/10.1103/PhysRevD.107.044022},
   DOI={10.1103/physrevd.107.044022},
   number={4},
   journal={Physical Review D},
   publisher={American Physical Society (APS)},
   author={Khyllep, Wompherdeiki and Dutta, Jibitesh and Saridakis, Emmanuel N. and Yesmakhanova, Kuralay},
   year={2023},
   month=feb }

@article{Holsclaw_2010,
   title={Nonparametric Dark Energy Reconstruction from Supernova Data},
   volume={105},
   ISSN={1079-7114},
   url={http://dx.doi.org/10.1103/PhysRevLett.105.241302},
   DOI={10.1103/physrevlett.105.241302},
   number={24},
   journal={Physical Review Letters},
   publisher={American Physical Society (APS)},
   author={Holsclaw, Tracy and Alam, Ujjaini and Sansó, Bruno and Lee, Herbert and Heitmann, Katrin and Habib, Salman and Higdon, David},
   year={2010},
   month=dec }

@article{Melia_2018,
   title={Model selection using cosmic chronometers with Gaussian Processes},
   volume={2018},
   ISSN={1475-7516},
   url={http://dx.doi.org/10.1088/1475-7516/2018/02/034},
   DOI={10.1088/1475-7516/2018/02/034},
   number={02},
   journal={Journal of Cosmology and Astroparticle Physics},
   publisher={IOP Publishing},
   author={Melia, Fulvio and Yennapureddy, Manoj K.},
   year={2018},
   month=feb, pages={034–034} }

@article{Pinho_2018,
   title={Model-independent reconstruction of the linear anisotropic stress η},
   volume={2018},
   ISSN={1475-7516},
   url={http://dx.doi.org/10.1088/1475-7516/2018/11/027},
   DOI={10.1088/1475-7516/2018/11/027},
   number={11},
   journal={Journal of Cosmology and Astroparticle Physics},
   publisher={IOP Publishing},
   author={Pinho, Ana Marta and Casas, Santiago and Amendola, Luca},
   year={2018},
   month=nov, pages={027–027} }

@article{Zhang_2018,
   title={Gaussian processes reconstruction of dark energy from observational data},
   volume={78},
   ISSN={1434-6052},
   url={http://dx.doi.org/10.1140/epjc/s10052-018-5953-3},
   DOI={10.1140/epjc/s10052-018-5953-3},
   number={6},
   journal={The European Physical Journal C},
   publisher={Springer Science and Business Media LLC},
   author={Zhang, Ming-Jian and Li, Hong},
   year={2018},
   month=jun }

@article{Yin_2019,
   title={Non-parametric reconstruction of growth index via Gaussian processes},
   volume={62},
   ISSN={1869-1927},
   url={http://dx.doi.org/10.1007/s11433-019-9373-0},
   DOI={10.1007/s11433-019-9373-0},
   number={9},
   journal={Science China Physics, Mechanics &amp; Astronomy},
   publisher={Springer Science and Business Media LLC},
   author={Yin, Zhao-Yu and Wei, Hao},
   year={2019},
   month=apr }

@article{Elizalde_2019,
   title={Swampland criteria for a dark energy dominated universe ensuing from Gaussian processes and 
<mml:math xmlns:mml="http://www.w3.org/1998/Math/MathML" display="inline"><mml:mi>H</mml:mi><mml:mo stretchy="false">(</mml:mo><mml:mi>z</mml:mi><mml:mo stretchy="false">)</mml:mo></mml:math>
 data analysis},
   volume={99},
   ISSN={2470-0029},
   url={http://dx.doi.org/10.1103/PhysRevD.99.103533},
   DOI={10.1103/physrevd.99.103533},
   number={10},
   journal={Physical Review D},
   publisher={American Physical Society (APS)},
   author={Elizalde, Emilio and Khurshudyan, Martiros},
   year={2019},
   month=may }

@article{Pinho_2018,
doi = {10.1088/1475-7516/2018/11/027},
url = {https://dx.doi.org/10.1088/1475-7516/2018/11/027},
year = {2018},
month = {nov},
publisher = {},
volume = {2018},
number = {11},
pages = {027},
author = {Ana Marta Pinho and Santiago Casas and Luca Amendola},
title = {Model-independent reconstruction of the linear anisotropic stress η},
journal = {Journal of Cosmology and Astroparticle Physics}
}
@article{Rau_2019,
   title={Estimating redshift distributions using hierarchical logistic Gaussian processes},
   volume={491},
   ISSN={1365-2966},
   url={http://dx.doi.org/10.1093/mnras/stz3295},
   DOI={10.1093/mnras/stz3295},
   number={4},
   journal={Monthly Notices of the Royal Astronomical Society},
   publisher={Oxford University Press (OUP)},
   author={Rau, Markus Michael and Wilson, Simon and Mandelbaum, Rachel},
   year={2019},
   month=nov, pages={4768–4782} }
@article{G_mez_Valent_2018,
   title={H0from cosmic chronometers and Type Ia supernovae, with Gaussian Processes and the novel Weighted Polynomial Regression method},
   volume={2018},
   ISSN={1475-7516},
   url={http://dx.doi.org/10.1088/1475-7516/2018/04/051},
   DOI={10.1088/1475-7516/2018/04/051},
   number={04},
   journal={Journal of Cosmology and Astroparticle Physics},
   publisher={IOP Publishing},
   author={Gómez-Valent, Adrià and Amendola, Luca},
   year={2018},
   month=apr, pages={051–051} }

@article{Busti_2014,
   title={Evidence for a lower value for H0 from cosmic chronometers data?},
   volume={441},
   ISSN={1745-3925},
   url={http://dx.doi.org/10.1093/mnrasl/slu035},
   DOI={10.1093/mnrasl/slu035},
   number={1},
   journal={Monthly Notices of the Royal Astronomical Society: Letters},
   publisher={Oxford University Press (OUP)},
   author={Busti, Vinicius C. and Clarkson, Chris and Seikel, Marina},
   year={2014},
   month=apr, pages={L11–L15} }


@misc{verde2014expansion,
      title={The expansion rate of the intermediate Universe in light of Planck}, 
      author={Licia Verde and Pavlos Protopapas and Raul Jimenez},
      year={2014},
      eprint={1403.2181},
      archivePrefix={arXiv},
      primaryClass={astro-ph.CO}
}

@article{Li_2016,
   title={Constructing a cosmological model-independent Hubble diagram of type Ia supernovae with cosmic chronometers},
   volume={93},
   ISSN={2470-0029},
   url={http://dx.doi.org/10.1103/PhysRevD.93.043014},
   DOI={10.1103/physrevd.93.043014},
   number={4},
   journal={Physical Review D},
   publisher={American Physical Society (APS)},
   author={Li, Zhengxiang and Gonzalez, J. E. and Yu, Hongwei and Zhu, Zong-Hong and Alcaniz, J. S.},
   year={2016},
   month=feb }

@article{Melia_2018,
   title={Model selection using cosmic chronometers with Gaussian Processes},
   volume={2018},
   ISSN={1475-7516},
   url={http://dx.doi.org/10.1088/1475-7516/2018/02/034},
   DOI={10.1088/1475-7516/2018/02/034},
   number={02},
   journal={Journal of Cosmology and Astroparticle Physics},
   publisher={IOP Publishing},
   author={Melia, Fulvio and Yennapureddy, Manoj K.},
   year={2018},
   month=feb, pages={034–034} }

@misc{wang2016modelindependent,
      title={Model-independent determination on $H_0$ using the latest $H(z)$ data}, 
      author={Deng Wang and Xin-He Meng},
      year={2016},
      eprint={1610.01202},
      archivePrefix={arXiv},
      primaryClass={gr-qc}
}





@article{Seikel_2012,
   title={Reconstruction of dark energy and expansion dynamics using Gaussian processes},
   volume={2012},
   ISSN={1475-7516},
   url={http://dx.doi.org/10.1088/1475-7516/2012/06/036},
   DOI={10.1088/1475-7516/2012/06/036},
   number={06},
   journal={Journal of Cosmology and Astroparticle Physics},
   publisher={IOP Publishing},
   author={Seikel, Marina and Clarkson, Chris and Smith, Mathew},
   year={2012},
   month=jun, pages={036–036} }

@article{Mehrabi_2021,
   title={Cosmographic Parameters in Model-independent Approaches},
   volume={923},
   ISSN={1538-4357},
   url={http://dx.doi.org/10.3847/1538-4357/ac2fff},
   DOI={10.3847/1538-4357/ac2fff},
   number={2},
   journal={The Astrophysical Journal},
   publisher={American Astronomical Society},
   author={Mehrabi, Ahmad and Rezaei, Mehdi},
   year={2021},
  month=dec, pages={274} }

@misc{exp2,
      title={Optimising Gaussian processes for reconstructing dark energy dynamics from supernovae}, 
      author={Marina Seikel and Chris Clarkson},
      year={2013},
      eprint={1311.6678},
      archivePrefix={arXiv},
      primaryClass={astro-ph.CO}
}
@article{Jimenez_2020,
   title={Cosmology in 
<mml:math xmlns:mml="http://www.w3.org/1998/Math/MathML" display="inline"><mml:mi>f</mml:mi><mml:mo stretchy="false">(</mml:mo><mml:mi>Q</mml:mi><mml:mo stretchy="false">)</mml:mo></mml:math>
 geometry},
   volume={101},
   ISSN={2470-0029},
   url={http://dx.doi.org/10.1103/PhysRevD.101.103507},
   DOI={10.1103/physrevd.101.103507},
   number={10},
   journal={Physical Review D},
   publisher={American Physical Society (APS)},
   author={Jiménez, Jose Beltrán and Heisenberg, Lavinia and Koivisto, Tomi and Pekar, Simon},
   year={2020},
   month=may }

@article{Lazkoz_2019,
   title={Observational constraints of 
<mml:math xmlns:mml="http://www.w3.org/1998/Math/MathML" display="inline"><mml:mi>f</mml:mi><mml:mo mathvariant="bold" stretchy="false">(</mml:mo><mml:mi>Q</mml:mi><mml:mo mathvariant="bold" stretchy="false">)</mml:mo></mml:math>
 gravity},
   volume={100},
   ISSN={2470-0029},
   url={http://dx.doi.org/10.1103/PhysRevD.100.104027},
   DOI={10.1103/physrevd.100.104027},
   number={10},
   journal={Physical Review D},
   publisher={American Physical Society (APS)},
   author={Lazkoz, Ruth and Lobo, Francisco S. N. and Ortiz-Baños, María and Salzano, Vincenzo},
   year={2019},
   month=nov }
@article{Khyllep_2023,
   title={Cosmology in 
<mml:math xmlns:mml="http://www.w3.org/1998/Math/MathML" display="inline"><mml:mi>f</mml:mi><mml:mo stretchy="false">(</mml:mo><mml:mi>Q</mml:mi><mml:mo stretchy="false">)</mml:mo></mml:math>
 gravity: A unified dynamical systems analysis of the background and perturbations},
   volume={107},
   ISSN={2470-0029},
   url={http://dx.doi.org/10.1103/PhysRevD.107.044022},
   DOI={10.1103/physrevd.107.044022},
   number={4},
   journal={Physical Review D},
   publisher={American Physical Society (APS)},
   author={Khyllep, Wompherdeiki and Dutta, Jibitesh and Saridakis, Emmanuel N. and Yesmakhanova, Kuralay},
   year={2023},
   month=feb }

@article{Anagnostopoulos_2023,
   title={New models and big bang nucleosynthesis constraints in f(Q) gravity},
   volume={83},
   ISSN={1434-6052},
   url={http://dx.doi.org/10.1140/epjc/s10052-023-11190-x},
   DOI={10.1140/epjc/s10052-023-11190-x},
   number={1},
   journal={The European Physical Journal C},
   publisher={Springer Science and Business Media LLC},
   author={Anagnostopoulos, Fotios K. and Gakis, Viktor and Saridakis, Emmanuel N. and Basilakos, Spyros},
   year={2023},
   month=jan }

@article{Sokoliuk_2023,
   title={On the impact off(Q) gravity on the large scale structure},
   volume={522},
   ISSN={1365-2966},
   url={http://dx.doi.org/10.1093/mnras/stad968},
   DOI={10.1093/mnras/stad968},
   number={1},
   journal={Monthly Notices of the Royal Astronomical Society},
   publisher={Oxford University Press (OUP)},
   author={Sokoliuk, Oleksii and Arora, Simran and Praharaj, Subhrat and Baransky, Alexander and Sahoo, P K},
   year={2023},
   month=mar, pages={252–267} }

@BOOK{exp1,
       author = {Rasmussen, Carl Edward and Williams, Christopher K. I.},
        title = "{Gaussian Processes for Machine Learning}",
    publisher = {The MIT Press},
	 year = "2005",
      edition = {1},
	 isbn = {9780262256834},
     url={ https://doi.org/10.7551/mitpress/3206.001.0001}
}
















@INPROCEEDINGS{1989BAAS...21..780H,
       author = {{Hanisch}, R.~J. and {Biemesderfer}, C.~D.},
        title = "{T$_{E}$X and LAT$_{E}$X Macro Definition Files for Astronomical Publications}",
    booktitle = {\baas},
         year = "1989",
        month = "Mar",
        pages = {780},
       adsurl = {https://ui.adsabs.harvard.edu/abs/1989BAAS...21..780H},
      adsnote = {Provided by the SAO/NASA Astrophysics Data System}
}

@BOOK{lamport94,
       author = {{Lamport}, L.},
        title = "{LaTeX: A Document Preparation System}",
    publisher = {Addison-Wesley Professional},
	 year = "1994",
      edition = {2},
	 isbn = {0201529831}
}



















@article{Daniel_2010,
   title={Cosmic chronometers: constraining the equation of state of dark energy. I: H(z) measurements},
   volume={2010},
   ISSN={1475-7516},
   url={http://dx.doi.org/10.1088/1475-7516/2010/02/008},
   DOI={10.1088/1475-7516/2010/02/008},
   number={02},
   journal={Journal of Cosmology and Astroparticle Physics},
   publisher={IOP Publishing},
   author={Daniel Stern and Raul Jimenez and Licia Verde and Marc Kamionkowski and S. Adam Stanford},
   year={2010},
   month=feb, pages={008–008} }

@article{Simon_2005,
   title={Constraints on the redshift dependence of the dark energy potential},
   volume={71},
   ISSN={1550-2368},
   url={http://dx.doi.org/10.1103/PhysRevD.71.123001},
   DOI={10.1103/physrevd.71.123001},
   number={12},
   journal={Physical Review D},
   publisher={American Physical Society (APS)},
   author={Simon, Joan and Verde, Licia and Jimenez, Raul},
   year={2005},
   month=jun }

@article{M_Moresco_2012,
   title={Improved constraints on the expansion rate of the Universe
 up to z ∼ 1.1 from the spectroscopic evolution of cosmic chronometers},
   volume={2012},
   ISSN={1475-7516},
   url={http://dx.doi.org/10.1088/1475-7516/2012/08/006},
   DOI={10.1088/1475-7516/2012/08/006},
   number={08},
   journal={Journal of Cosmology and Astroparticle Physics},
   publisher={IOP Publishing},
   author={M. Moresco and A. Cimatti and R. Jimenez and L. Pozzetti and G. Zamorani and M. Bolzonella and J. Dunlop and F. Lamareille and M. Mignoli and H. Pearce and P. Rosati and D. Stern and L. Verde and E. Zucca and C.M. Carollo and T. Contini and J.-P. Kneib and O. Le Fèvre and S.J. Lilly and V. Mainieri and A. Renzini and M. Scodeggio and I. Balestra and R. Gobat and R. McLure and S. Bardelli and A. Bongiorno and K. Caputi and O. Cucciati and S. de la Torre and L. de Ravel and P. Franzetti and B. Garilli and A. Iovino and P. Kampczyk and C. Knobel and K. Kovač and J.-F. Le Borgne and V. Le Brun and C. Maier and R. Pelló and Y. Peng and E. Perez-Montero and V. Presotto and J.D. Silverman and M. Tanaka and L.A.M. Tasca and L. Tresse and D. Vergani and O. Almaini and L. Barnes and R. Bordoloi and E. Bradshaw and A. Cappi and R. Chuter and M. Cirasuolo and G. Coppa and C. Diener and S. Foucaud and W. Hartley and M. Kamionkowski and A.M. Koekemoer and C. López-Sanjuan and H.J. McCracken and P. Nair and P. Oesch and A. Stanford and N. Welikala},
   year={2012},
   month=aug, pages={006–006} }

@article{Zhang_2014,
doi = {10.1088/1674-4527/14/10/002},
url = {https://dx.doi.org/10.1088/1674-4527/14/10/002},
year = {2014},
month = {oct},
publisher = {},
volume = {14},
number = {10},
pages = {1221},
author = { Cong, Zhang and Han, Zhang and Shuo, Yuan and Siqi, Liu and TongJie, Zhang and Yan-Chun, Sun},
title = {Four new observational H(z) data from luminous red galaxies in the Sloan Digital Sky Survey data release seven},
journal = {Research in Astronomy and Astrophysics}
}

@article{Moresco_2016,
   title={A 6
   volume={2016},
   ISSN={1475-7516},
   url={http://dx.doi.org/10.1088/1475-7516/2016/05/014},
   DOI={10.1088/1475-7516/2016/05/014},
   number={05},
   journal={Journal of Cosmology and Astroparticle Physics},
   publisher={IOP Publishing},
   author={Moresco, Michele and Pozzetti, Lucia and Cimatti, Andrea and Jimenez, Raul and Maraston, Claudia and Verde, Licia and Thomas, Daniel and Citro, Annalisa and Tojeiro, Rita and Wilkinson, David},
   year={2016},
   month=may, pages={014–014} }

@article{Ratsimbazafy_2017,
   title={Age-dating luminous red galaxies observed with the Southern African Large Telescope},
   volume={467},
   ISSN={1365-2966},
   url={http://dx.doi.org/10.1093/mnras/stx301},
   DOI={10.1093/mnras/stx301},
   number={3},
   journal={Monthly Notices of the Royal Astronomical Society},
   publisher={Oxford University Press (OUP)},
   author={Ratsimbazafy, A. L. and Loubser, S. I. and Crawford, S. M. and Cress, C. M. and Bassett, B. A. and Nichol, R. C. and Väisänen, P.},
   year={2017},
   month=feb, pages={3239–3254} }

@article{Moresco_2015,
   title={Raising the bar: new constraints on the Hubble parameter with cosmic chronometers at z ∼ 2},
   volume={450},
   ISSN={1745-3925},
   url={http://dx.doi.org/10.1093/mnrasl/slv037},
   DOI={10.1093/mnrasl/slv037},
   number={1},
   journal={Monthly Notices of the Royal Astronomical Society: Letters},
   publisher={Oxford University Press (OUP)},
   author={Moresco, Michele},
   year={2015},
   month=apr, pages={L16–L20} }

@article{Gaztaga_2009,
   title={Clustering of luminous red galaxies - IV. Baryon acoustic peak in the line-of-sight direction and a direct measurement ofH(z)},
   volume={399},
   ISSN={1365-2966},
   url={http://dx.doi.org/10.1111/j.1365-2966.2009.15405.x},
   DOI={10.1111/j.1365-2966.2009.15405.x},
   number={3},
   journal={Monthly Notices of the Royal Astronomical Society},
   publisher={Oxford University Press (OUP)},
   author={Gaztañaga, Enrique and Cabré, Anna and Hui, Lam},
   year={2009},
   month=nov, pages={1663–1680} }

@article{Oka_2014,
   title={Simultaneous constraints on the growth of structure and cosmic expansion from the multipole power spectra of the SDSS DR7 LRG sample},
   volume={439},
   ISSN={1365-2966},
   url={http://dx.doi.org/10.1093/mnras/stu111},
   DOI={10.1093/mnras/stu111},
   number={3},
   journal={Monthly Notices of the Royal Astronomical Society},
   publisher={Oxford University Press (OUP)},
   author={Oka, A. and Saito, S. and Nishimichi, T. and Taruya, A. and Yamamoto, K.},
   year={2014},
   month=feb, pages={2515–2530} }

@article{Wang_2017,
   title={The clustering of galaxies in the completed SDSS-III Baryon Oscillation Spectroscopic Survey: tomographic BAO analysis of DR12 combined sample in configuration space},
   volume={469},
   ISSN={1365-2966},
   url={http://dx.doi.org/10.1093/mnras/stx1090},
   DOI={10.1093/mnras/stx1090},
   number={3},
   journal={Monthly Notices of the Royal Astronomical Society},
   publisher={Oxford University Press (OUP)},
   author={Wang, Yuting and Zhao, Gong-Bo and Chuang, Chia-Hsun and Ross, Ashley J. and Percival, Will J. and Gil-Marín, Héctor and Cuesta, Antonio J. and Kitaura, Francisco-Shu and Rodriguez-Torres, Sergio and Brownstein, Joel R. and Eisenstein, Daniel J. and Ho, Shirley and Kneib, Jean-Paul and Olmstead, Matthew D. and Prada, Francisco and Rossi, Graziano and Sánchez, Ariel G. and Salazar-Albornoz, Salvador and Thomas, Daniel and Tinker, Jeremy and Tojeiro, Rita and Vargas-Magaña, Mariana and Zhu, Fangzhou},
   year={2017},
   month=may, pages={3762–3774} }

@article{Chuang_2013,
   title={Modelling the anisotropic two-point galaxy correlation function on small scales and single-probe measurements of H(z), DA(z) and f(z) 8(z) from the Sloan Digital Sky Survey DR7 luminous red galaxies},
   volume={435},
   ISSN={1365-2966},
   url={http://dx.doi.org/10.1093/mnras/stt1290},
   DOI={10.1093/mnras/stt1290},
   number={1},
   journal={Monthly Notices of the Royal Astronomical Society},
   publisher={Oxford University Press (OUP)},
   author={Chuang, C.-H. and Wang, Y.},
   year={2013},
   month=aug, pages={255–262} }

@article{Alam_2017,
   title={The clustering of galaxies in the completed SDSS-III Baryon Oscillation Spectroscopic Survey: cosmological analysis of the DR12 galaxy sample},
   volume={470},
   ISSN={1365-2966},
   url={http://dx.doi.org/10.1093/mnras/stx721},
   DOI={10.1093/mnras/stx721},
   number={3},
   journal={Monthly Notices of the Royal Astronomical Society},
   publisher={Oxford University Press (OUP)},
   author={Alam, Shadab and Ata, Metin and Bailey, Stephen and Beutler, Florian and Bizyaev, Dmitry and Blazek, Jonathan A. and Bolton, Adam S. and Brownstein, Joel R. and Burden, Angela and Chuang, Chia-Hsun and Comparat, Johan and Cuesta, Antonio J. and Dawson, Kyle S. and Eisenstein, Daniel J. and Escoffier, Stephanie and Gil-Marín, Héctor and Grieb, Jan Niklas and Hand, Nick and Ho, Shirley and Kinemuchi, Karen and Kirkby, David and Kitaura, Francisco and Malanushenko, Elena and Malanushenko, Viktor and Maraston, Claudia and McBride, Cameron K. and Nichol, Robert C. and Olmstead, Matthew D. and Oravetz, Daniel and Padmanabhan, Nikhil and Palanque-Delabrouille, Nathalie and Pan, Kaike and Pellejero-Ibanez, Marcos and Percival, Will J. and Petitjean, Patrick and Prada, Francisco and Price-Whelan, Adrian M. and Reid, Beth A. and Rodríguez-Torres, Sergio A. and Roe, Natalie A. and Ross, Ashley J. and Ross, Nicholas P. and Rossi, Graziano and Rubiño-Martín, Jose Alberto and Saito, Shun and Salazar-Albornoz, Salvador and Samushia, Lado and Sánchez, Ariel G. and Satpathy, Siddharth and Schlegel, David J. and Schneider, Donald P. and Scóccola, Claudia G. and Seo, Hee-Jong and Sheldon, Erin S. and Simmons, Audrey and Slosar, Anže and Strauss, Michael A. and Swanson, Molly E. C. and Thomas, Daniel and Tinker, Jeremy L. and Tojeiro, Rita and Magaña, Mariana Vargas and Vazquez, Jose Alberto and Verde, Licia and Wake, David A. and Wang, Yuting and Weinberg, David H. and White, Martin and Wood-Vasey, W. Michael and Yèche, Christophe and Zehavi, Idit and Zhai, Zhongxu and Zhao, Gong-Bo},
   year={2017},
   month=mar, pages={2617–2652} }

@article{Blake_2012,
   title={The WiggleZ Dark Energy Survey: joint measurements of the expansion and growth history atz&lt; 1: WiggleZ Survey: expansion history},
   volume={425},
   ISSN={0035-8711},
   url={http://dx.doi.org/10.1111/j.1365-2966.2012.21473.x},
   DOI={10.1111/j.1365-2966.2012.21473.x},
   number={1},
   journal={Monthly Notices of the Royal Astronomical Society},
   publisher={Oxford University Press (OUP)},
   author={Blake, Chris and Brough, Sarah and Colless, Matthew and Contreras, Carlos and Couch, Warrick and Croom, Scott and Croton, Darren and Davis, Tamara M. and Drinkwater, Michael J. and Forster, Karl and Gilbank, David and Gladders, Mike and Glazebrook, Karl and Jelliffe, Ben and Jurek, Russell J. and Li, I-hui and Madore, Barry and Martin, D. Christopher and Pimbblet, Kevin and Poole, Gregory B. and Pracy, Michael and Sharp, Rob and Wisnioski, Emily and Woods, David and Wyder, Ted K. and Yee, H. K. C.},
   year={2012},
   month=jul, pages={405–414} }

@article{Chuang_2013,
    author = {Chuang, Chia-Hsun and Prada, Francisco and Cuesta, Antonio J. and Eisenstein, Daniel J. and Kazin, Eyal and Padmanabhan, Nikhil and Sánchez, Ariel G. and Xu, Xiaoying and Beutler, Florian and Manera, Marc and Schlegel, David J and Schneider, Donald P. and Weinberg, David H. and Brinkmann, Jon and Brownstein, Joel R. and Thomas, Daniel},
    title = "{The clustering of galaxies in the SDSS-III Baryon Oscillation Spectroscopic Survey: single-probe measurements and the strong power of f(z)σ8(z) on constraining dark energy}",
    journal = {Monthly Notices of the Royal Astronomical Society},
    volume = {433},
    number = {4},
    pages = {3559-3571},
    year = {2013},
    month = {07},
    issn = {0035-8711},
    doi = {10.1093/mnras/stt988},
    url = {https://doi.org/10.1093/mnras/stt988},
    eprint = {https://academic.oup.com/mnras/article-pdf/433/4/3559/4931552/stt988.pdf},
}

@article{Anderson_2014,
   title={The clustering of galaxies in the SDSS-III Baryon Oscillation Spectroscopic Survey: baryon acoustic oscillations in the Data Releases 10 and 11 Galaxy samples},
   volume={441},
   ISSN={0035-8711},
   url={http://dx.doi.org/10.1093/mnras/stu523},
   DOI={10.1093/mnras/stu523},
   number={1},
   journal={Monthly Notices of the Royal Astronomical Society},
   publisher={Oxford University Press (OUP)},
   author={Anderson, Lauren and Aubourg, Éric and Bailey, Stephen and Beutler, Florian and Bhardwaj, Vaishali and Blanton, Michael and Bolton, Adam S. and Brinkmann, J. and Brownstein, Joel R. and Burden, Angela and Chuang, Chia-Hsun and Cuesta, Antonio J. and Dawson, Kyle S. and Eisenstein, Daniel J. and Escoffier, Stephanie and Gunn, James E. and Guo, Hong and Ho, Shirley and Honscheid, Klaus and Howlett, Cullan and Kirkby, David and Lupton, Robert H. and Manera, Marc and Maraston, Claudia and McBride, Cameron K. and Mena, Olga and Montesano, Francesco and Nichol, Robert C. and Nuza, Sebastián E. and Olmstead, Matthew D. and Padmanabhan, Nikhil and Palanque-Delabrouille, Nathalie and Parejko, John and Percival, Will J. and Petitjean, Patrick and Prada, Francisco and Price-Whelan, Adrian M. and Reid, Beth and Roe, Natalie A. and Ross, Ashley J. and Ross, Nicholas P. and Sabiu, Cristiano G. and Saito, Shun and Samushia, Lado and Sánchez, Ariel G. and Schlegel, David J. and Schneider, Donald P. and Scoccola, Claudia G. and Seo, Hee-Jong and Skibba, Ramin A. and Strauss, Michael A. and Swanson, Molly E. C. and Thomas, Daniel and Tinker, Jeremy L. and Tojeiro, Rita and Magaña, Mariana Vargas and Verde, Licia and Wake, David A. and Weaver, Benjamin A. and Weinberg, David H. and White, Martin and Xu, Xiaoying and Yèche, Christophe and Zehavi, Idit and Zhao, Gong-Bo},
   year={2014},
   month=apr, pages={24–62} }

@article{Busca_2013,
   title={Baryon acoustic oscillations in the Lyαforest of BOSS quasars},
   volume={552},
   ISSN={1432-0746},
   url={http://dx.doi.org/10.1051/0004-6361/201220724},
   DOI={10.1051/0004-6361/201220724},
   journal={Astronomy &amp; Astrophysics},
   publisher={EDP Sciences},
   author={Busca, N. G. and Delubac, T. and Rich, J. and Bailey, S. and Font-Ribera, A. and Kirkby, D. and Le Goff, J.-M. and Pieri, M. M. and Slosar, A. and Aubourg, É. and Bautista, J. E. and Bizyaev, D. and Blomqvist, M. and Bolton, A. S. and Bovy, J. and Brewington, H. and Borde, A. and Brinkmann, J. and Carithers, B. and Croft, R. A. C. and Dawson, K. S. and Ebelke, G. and Eisenstein, D. J. and Hamilton, J.-C. and Ho, S. and Hogg, D. W. and Honscheid, K. and Lee, K.-G. and Lundgren, B. and Malanushenko, E. and Malanushenko, V. and Margala, D. and Maraston, C. and Mehta, K. and Miralda-Escudé, J. and Myers, A. D. and Nichol, R. C. and Noterdaeme, P. and Olmstead, M. D. and Oravetz, D. and Palanque-Delabrouille, N. and Pan, K. and Pâris, I. and Percival, W. J. and Petitjean, P. and Roe, N. A. and Rollinde, E. and Ross, N. P. and Rossi, G. and Schlegel, D. J. and Schneider, D. P. and Shelden, A. and Sheldon, E. S. and Simmons, A. and Snedden, S. and Tinker, J. L. and Viel, M. and Weaver, B. A. and Weinberg, D. H. and White, M. and Yèche, C. and York, D. G.},
   year={2013},
   month=apr, pages={A96} }




@article{Bautista_2017,
   title={Measurement of baryon acoustic oscillation correlations atz=2.3 with SDSS DR12 Lyα-Forests},
   volume={603},
   ISSN={1432-0746},
   url={http://dx.doi.org/10.1051/0004-6361/201730533},
   DOI={10.1051/0004-6361/201730533},
   journal={Astronomy &amp; Astrophysics},
   publisher={EDP Sciences},
   author={Bautista, Julian E. and Busca, Nicolás G. and Guy, Julien and Rich, James and Blomqvist, Michael and du Mas des Bourboux, Hélion and Pieri, Matthew M. and Font-Ribera, Andreu and Bailey, Stephen and Delubac, Timothée and Kirkby, David and Le Goff, Jean-Marc and Margala, Daniel and Slosar, Anže and Vazquez, Jose Alberto and Brownstein, Joel R. and Dawson, Kyle S. and Eisenstein, Daniel J. and Miralda-Escudé, Jordi and Noterdaeme, Pasquier and Palanque-Delabrouille, Nathalie and Pâris, Isabelle and Petitjean, Patrick and Ross, Nicholas P. and Schneider, Donald P. and Weinberg, David H. and Yèche, Christophe},
   year={2017},
   month=jun, pages={A12} }


@article{Delubac_2015,
   title={Baryon acoustic oscillations in the Lyαforest of BOSS DR11 quasars},
   volume={574},
   ISSN={1432-0746},
   url={http://dx.doi.org/10.1051/0004-6361/201423969},
   DOI={10.1051/0004-6361/201423969},
   journal={Astronomy &amp; Astrophysics},
   publisher={EDP Sciences},
   author={Delubac, Timothée and Bautista, Julian E. and Busca, Nicolás G. and Rich, James and Kirkby, David and Bailey, Stephen and Font-Ribera, Andreu and Slosar, Anže and Lee, Khee-Gan and Pieri, Matthew M. and Hamilton, Jean-Christophe and Aubourg, Éric and Blomqvist, Michael and Bovy, Jo and Brinkmann, Jon and Carithers, William and Dawson, Kyle S. and Eisenstein, Daniel J. and Gontcho A Gontcho, Satya and Kneib, Jean-Paul and Le Goff, Jean-Marc and Margala, Daniel and Miralda-Escudé, Jordi and Myers, Adam D. and Nichol, Robert C. and Noterdaeme, Pasquier and O’Connell, Ross and Olmstead, Matthew D. and Palanque-Delabrouille, Nathalie and Pâris, Isabelle and Petitjean, Patrick and Ross, Nicholas P. and Rossi, Graziano and Schlegel, David J. and Schneider, Donald P. and Weinberg, David H. and Yèche, Christophe and York, Donald G.},
   year={2015},
   month=jan, pages={A59} }

@article{Font_Ribera_2014,
   title={Quasar-Lyman α forest cross-correlation from BOSS DR11: Baryon Acoustic Oscillations},
   volume={2014},
   ISSN={1475-7516},
   url={http://dx.doi.org/10.1088/1475-7516/2014/05/027},
   DOI={10.1088/1475-7516/2014/05/027},
   number={05},
   journal={Journal of Cosmology and Astroparticle Physics},
   publisher={IOP Publishing},
   author={Font-Ribera, Andreu and Kirkby, David and Busca, Nicolas and Miralda-Escudé, Jordi and Ross, Nicholas P. and Slosar, Anže and Rich, James and Aubourg, Éric and Bailey, Stephen and Bhardwaj, Vaishali and Bautista, Julian and Beutler, Florian and Bizyaev, Dmitry and Blomqvist, Michael and Brewington, Howard and Brinkmann, Jon and Brownstein, Joel R. and Carithers, Bill and Dawson, Kyle S. and Delubac, Timothée and Ebelke, Garrett and Eisenstein, Daniel J. and Ge, Jian and Kinemuchi, Karen and Lee, Khee-Gan and Malanushenko, Viktor and Malanushenko, Elena and Marchante, Moses and Margala, Daniel and Muna, Demitri and Myers, Adam D. and Noterdaeme, Pasquier and Oravetz, Daniel and Palanque-Delabrouille, Nathalie and Pâris, Isabelle and Petitjean, Patrick and Pieri, Matthew M. and Rossi, Graziano and Schneider, Donald P. and Simmons, Audrey and Viel, Matteo and Yeche, Christophe and York, Donald G.},
   year={2014},
   month=may, pages={027–027} }

@article{Cai_2020N,
   title={Model-independent Reconstruction of f(T) Gravity from Gaussian Processes},
   volume={888},
   ISSN={1538-4357},
   url={http://dx.doi.org/10.3847/1538-4357/ab5a7f},
   DOI={10.3847/1538-4357/ab5a7f},
   number={2},
   journal={The Astrophysical Journal},
   publisher={American Astronomical Society},
   author={Cai, Yi-Fu and Khurshudyan, Martiros and Saridakis, Emmanuel N.},
   year={2020},
   month=jan, pages={62} }

@article{Ren_2022N,
   title={Gaussian Processes and Effective Field Theory of f(T) Gravity under the H
               0 Tension},
   volume={932},
   ISSN={1538-4357},
   url={http://dx.doi.org/10.3847/1538-4357/ac6ba5},
   DOI={10.3847/1538-4357/ac6ba5},
   number={2},
   journal={The Astrophysical Journal},
   publisher={American Astronomical Society},
   author={Ren, Xin and Yan, Sheng-Feng and Zhao, Yaqi and Cai, Yi-Fu and Saridakis, Emmanuel N.},
   year={2022},
   month=jun, pages={131} }

@article{Borghi_2022,
   title={Toward a Better Understanding of Cosmic Chronometers: A New Measurement of H(z) at z ∼ 0.7},
   volume={928},
   ISSN={2041-8213},
   url={http://dx.doi.org/10.3847/2041-8213/ac3fb2},
   DOI={10.3847/2041-8213/ac3fb2},
   number={1},
   journal={The Astrophysical Journal Letters},
   publisher={American Astronomical Society},
   author={Borghi, Nicola and Moresco, Michele and Cimatti, Andrea},
   year={2022},
   month=mar, pages={L4} }

@article{Fortunato_2024,
   title={Search for the f(R, T) gravity functional form via gaussian processes},
   volume={84},
   ISSN={1434-6052},
   url={http://dx.doi.org/10.1140/epjc/s10052-024-12544-9},
   DOI={10.1140/epjc/s10052-024-12544-9},
   number={2},
   journal={The European Physical Journal C},
   publisher={Springer Science and Business Media LLC},
   author={Fortunato, J. A. S. and Moraes, P. H. R. S. and de Lima Júnior, J. G. and Brito, E.},
   year={2024},
   month=feb }


\begin{thebibliography}{}
\expandafter\ifx\csname natexlab\endcsname\relax\def\natexlab#1{#1}\fi
\providecommand{\url}[1]{\href{#1}{#1}}
\providecommand{\dodoi}[1]{doi:~\href{http://doi.org/#1}{\nolinkurl{#1}}}
\providecommand{\doeprint}[1]{\href{http://ascl.net/#1}{\nolinkurl{http://ascl.net/#1}}}
\providecommand{\doarXiv}[1]{\href{https://arxiv.org/abs/#1}{\nolinkurl{https://arxiv.org/abs/#1}}}

\bibitem[{Alam {et~al.}(2017)Alam, Ata, Bailey, Beutler, Bizyaev, Blazek, Bolton, Brownstein, Burden, Chuang, Comparat, Cuesta, Dawson, Eisenstein, Escoffier, Gil-Marín, Grieb, Hand, Ho, Kinemuchi, Kirkby, Kitaura, Malanushenko, Malanushenko, Maraston, McBride, Nichol, Olmstead, Oravetz, Padmanabhan, Palanque-Delabrouille, Pan, Pellejero-Ibanez, Percival, Petitjean, Prada, Price-Whelan, Reid, Rodríguez-Torres, Roe, Ross, Ross, Rossi, Rubiño-Martín, Saito, Salazar-Albornoz, Samushia, Sánchez, Satpathy, Schlegel, Schneider, Scóccola, Seo, Sheldon, Simmons, Slosar, Strauss, Swanson, Thomas, Tinker, Tojeiro, Magaña, Vazquez, Verde, Wake, Wang, Weinberg, White, Wood-Vasey, Yèche, Zehavi, Zhai, \& Zhao}]{Alam_2017}
Alam, S., Ata, M., Bailey, S., {et~al.} 2017, Monthly Notices of the Royal Astronomical Society, 470, 2617–2652, \dodoi{10.1093/mnras/stx721}

\bibitem[{Anagnostopoulos {et~al.}(2021)Anagnostopoulos, Basilakos, \& Saridakis}]{Anagnostopoulos_2021}
Anagnostopoulos, F.~K., Basilakos, S., \& Saridakis, E.~N. 2021, Physics Letters B, 822, 136634, \dodoi{10.1016/j.physletb.2021.136634}

\bibitem[{Anagnostopoulos {et~al.}(2023)Anagnostopoulos, Gakis, Saridakis, \& Basilakos}]{Anagnostopoulos_2023}
Anagnostopoulos, F.~K., Gakis, V., Saridakis, E.~N., \& Basilakos, S. 2023, The European Physical Journal C, 83, \dodoi{10.1140/epjc/s10052-023-11190-x}

\bibitem[{Anderson {et~al.}(2014)Anderson, Aubourg, Bailey, Beutler, Bhardwaj, Blanton, Bolton, Brinkmann, Brownstein, Burden, Chuang, Cuesta, Dawson, Eisenstein, Escoffier, Gunn, Guo, Ho, Honscheid, Howlett, Kirkby, Lupton, Manera, Maraston, McBride, Mena, Montesano, Nichol, Nuza, Olmstead, Padmanabhan, Palanque-Delabrouille, Parejko, Percival, Petitjean, Prada, Price-Whelan, Reid, Roe, Ross, Ross, Sabiu, Saito, Samushia, Sánchez, Schlegel, Schneider, Scoccola, Seo, Skibba, Strauss, Swanson, Thomas, Tinker, Tojeiro, Magaña, Verde, Wake, Weaver, Weinberg, White, Xu, Yèche, Zehavi, \& Zhao}]{Anderson_2014}
Anderson, L., Aubourg,  ., Bailey, S., {et~al.} 2014, Monthly Notices of the Royal Astronomical Society, 441, 24–62, \dodoi{10.1093/mnras/stu523}

\bibitem[{Ayuso {et~al.}(2021)Ayuso, Lazkoz, \& Salzano}]{Ayuso_2021}
Ayuso, I., Lazkoz, R., \& Salzano, V. 2021, Physical Review D, 103, \dodoi{10.1103/physrevd.103.063505}

\bibitem[{Bautista {et~al.}(2017)Bautista, Busca, Guy, Rich, Blomqvist, du~Mas~des Bourboux, Pieri, Font-Ribera, Bailey, Delubac, Kirkby, Le~Goff, Margala, Slosar, Vazquez, Brownstein, Dawson, Eisenstein, Miralda-Escudé, Noterdaeme, Palanque-Delabrouille, Pâris, Petitjean, Ross, Schneider, Weinberg, \& Yèche}]{Bautista_2017}
Bautista, J.E., Busca, N.G., Guy, J., {et al.} 2017, Astronomy and Astrophysics, 603, A12, \dodoi{10.1051/0004-6361/201730533}

\bibitem[{Blake {et~al.}(2012)Blake, Brough, Colless, Contreras, Couch, Croom, Croton, Davis, Drinkwater, Forster, Gilbank, Gladders, Glazebrook, Jelliffe, Jurek, Li, Madore, Martin, Pimbblet, Poole, Pracy, Sharp, Wisnioski, Woods, Wyder, \& Yee}]{Blake_2012}
Blake, C., Brough, S., Colless, M., {et~al.} 2012, Monthly Notices of the Royal Astronomical Society, 425, 405–414, \dodoi{10.1111/j.1365-2966.2012.21473.x}

\bibitem[{Borghi {et~al.}(2022)Borghi, Moresco, \& Cimatti}]{Borghi_2022}
Borghi, N., Moresco, M., \& Cimatti, A. 2022, The Astrophysical Journal Letters, 928, L4, \dodoi{10.3847/2041-8213/ac3fb2}

\bibitem[{Busca {et~al.}(2013)Busca, Delubac, Rich, Bailey, Font-Ribera, Kirkby, Le~Goff, Pieri, Slosar, Aubourg, Bautista, Bizyaev, Blomqvist, Bolton, Bovy, Brewington, Borde, Brinkmann, Carithers, Croft, Dawson, Ebelke, Eisenstein, Hamilton, Ho, Hogg, Honscheid, Lee, Lundgren, Malanushenko, Malanushenko, Margala, Maraston, Mehta, Miralda-Escudé, Myers, Nichol, Noterdaeme, Olmstead, Oravetz, Palanque-Delabrouille, Pan, Pâris, Percival, Petitjean, Roe, Rollinde, Ross, Rossi, Schlegel, Schneider, Shelden, Sheldon, Simmons, Snedden, Tinker, Viel, Weaver, Weinberg, White, Yèche, \& York}]{Busca_2013}
Busca, N.G., Delubac, T., Rich, J., {et al.} 2013, Astronomy and Astrophysics, 552, A96, \dodoi{10.1051/0004-6361/201220724}

\bibitem[{Busti {et~al.}(2014)Busti, Clarkson, \& Seikel}]{Busti_2014}
Busti, V.~C., Clarkson, C., \& Seikel, M. 2014, Monthly Notices of the Royal Astronomical Society: Letters, 441, L11–L15, \dodoi{10.1093/mnrasl/slu035}

\bibitem[{Cai {et~al.}(2016)Cai, Capozziello, De~Laurentis, \& Saridakis}]{Cai_2016}
Cai, Y.-F., Capozziello, S., De~Laurentis, M., \& Saridakis, E.~N. 2016, Reports on Progress in Physics, 79, 106901, \dodoi{10.1088/0034-4885/79/10/106901}

\bibitem[{Cai {et~al.}(2020)Cai, Khurshudyan, \& Saridakis}]{Cai_2020N}
Cai, Y.-F., Khurshudyan, M., \& Saridakis, E.~N. 2020, The Astrophysical Journal, 888, 62, \dodoi{10.3847/1538-4357/ab5a7f}

\bibitem[{Cai {et~al.}(2010)Cai, Saridakis, Setare, \& Xia}]{Cai_2010}
Cai, Y.-F., Saridakis, E.~N., Setare, M.~R., \& Xia, J.-Q. 2010, Physics Reports, 493, 1–60, \dodoi{10.1016/j.physrep.2010.04.001}

\bibitem[{Capozziello {et~al.}(2023)Capozziello, De~Falco, \& Ferrara}]{Capozziello_2023}
Capozziello, S., De~Falco, V., \& Ferrara, C. 2023, The European Physical Journal C, 83, \dodoi{10.1140/epjc/s10052-023-12072-y}

\bibitem[{Capozziello \& De~Laurentis(2011)}]{Capozziello_2011}
Capozziello, S., \& De~Laurentis, M. 2011, Physics Reports, 509, 167–321, \dodoi{10.1016/j.physrep.2011.09.003}

\bibitem[{Chuang \& Wang(2013)}]{Chuang_2013}
Chuang, C.-H., \& Wang, Y. 2013, Monthly Notices of the Royal Astronomical Society, 435, 255–262, \dodoi{10.1093/mnras/stt1290}

\bibitem[{Cong {et~al.}(2014)Cong, Han, Shuo, Siqi, TongJie, \& Yan-Chun}]{Zhang_2014}
Cong, Z., Han, Z., Shuo, Y., {et~al.} 2014, Research in Astronomy and Astrophysics, 14, 1221, \dodoi{10.1088/1674-4527/14/10/002}

\bibitem[{De {et~al.}(2024)De, Loo, \& Saridakis}]{De_2024}
De, A., Loo, T.-H., \& Saridakis, E.~N. 2024, Journal of Cosmology and Astroparticle Physics, 2024, 050, \dodoi{10.1088/1475-7516/2024/03/050}

\bibitem[{Delubac {et~al.}(2015)Delubac, Bautista, Busca, Rich, Kirkby, Bailey, Font-Ribera, Slosar, Lee, Pieri, Hamilton, Aubourg, Blomqvist, Bovy, Brinkmann, Carithers, Dawson, Eisenstein, Gontcho A~Gontcho, Kneib, Le~Goff, Margala, Miralda-Escudé, Myers, Nichol, Noterdaeme, O’Connell, Olmstead, Palanque-Delabrouille, Pâris, Petitjean, Ross, Rossi, Schlegel, Schneider, Weinberg, Yèche, \& York}]{Delubac_2015}
Delubac, T., Bautista, J.E., Busca, N.G., {et al.} 2015, Astronomy and Astrophysics, 574, A59, \dodoi{10.1051/0004-6361/201423969}

\bibitem[{Elizalde {et~al.}(2019)Elizalde, Khurshudyan, \& Nojiri}]{Elizalde_2019}
Elizalde, E., Khurshudyan, M., \& Nojiri, S. 2019, International Journal of Modern Physics D, 28, 1950019, \dodoi{10.1142/s0218271819500196}

\bibitem[{Font-Ribera {et~al.}(2014)Font-Ribera, Kirkby, Busca, Miralda-Escudé, Ross, Slosar, Rich, Aubourg, Bailey, Bhardwaj, Bautista, Beutler, Bizyaev, Blomqvist, Brewington, Brinkmann, Brownstein, Carithers, Dawson, Delubac, Ebelke, Eisenstein, Ge, Kinemuchi, Lee, Malanushenko, Malanushenko, Marchante, Margala, Muna, Myers, Noterdaeme, Oravetz, Palanque-Delabrouille, Pâris, Petitjean, Pieri, Rossi, Schneider, Simmons, Viel, Yeche, \& York}]{Font_Ribera_2014}
Font-Ribera, A., Kirkby, D., Busca, N., {et~al.} 2014, Journal of Cosmology and Astroparticle Physics, 2014, 027–027, \dodoi{10.1088/1475-7516/2014/05/027}

\bibitem[{Fortunato {et~al.}(2024)Fortunato, Moraes, de~Lima~Júnior, \& Brito}]{Fortunato_2024}
Fortunato, J. A.~S., Moraes, P. H. R.~S., de~Lima~Júnior, J.~G., \& Brito, E. 2024, The European Physical Journal C, 84, \dodoi{10.1140/epjc/s10052-024-12544-9}

\bibitem[{Freedman {et~al.}(2019)Freedman, Madore, Hatt, Hoyt, Jang, Beaton, Burns, Lee, Monson, Neeley, Phillips, Rich, \& Seibert}]{Freedman_2019}
Freedman, W.~L., Madore, B.~F., Hatt, D., {et~al.} 2019, The Astrophysical Journal, 882, 34, \dodoi{10.3847/1538-4357/ab2f73}

\bibitem[{Frusciante(2021)}]{Frusciante_2021}
Frusciante, N. 2021, Physical Review D, 103, \dodoi{10.1103/physrevd.103.044021}

\bibitem[{Gaztañaga {et~al.}(2009)Gaztañaga, Cabré, \& Hui}]{Gaztaga_2009}
Gaztañaga, E., Cabré, A., \& Hui, L. 2009, Monthly Notices of the Royal Astronomical Society, 399, 1663–1680, \dodoi{10.1111/j.1365-2966.2009.15405.x}

\bibitem[{Gómez-Valent \& Amendola(2018)}]{G_mez_Valent_2018}
Gómez-Valent, A., \& Amendola, L. 2018, Journal of Cosmology and Astroparticle Physics, 2018, 051–051, \dodoi{10.1088/1475-7516/2018/04/051}

\bibitem[{Harko {et~al.}(2018)Harko, Koivisto, Lobo, Olmo, \& Rubiera-Garcia}]{Harko_2018}
Harko, T., Koivisto, T.~S., Lobo, F.~S., Olmo, G.~J., \& Rubiera-Garcia, D. 2018, Physical Review D, 98, \dodoi{10.1103/physrevd.98.084043}

\bibitem[{Heisenberg(2023)}]{heisenberg2023review}
Heisenberg, L. 2023, Review on $f(Q)$ Gravity.
\newblock \url{https://doi.org/10.1016/j.physrep.2024.02.001}

\bibitem[{Holsclaw {et~al.}(2010)Holsclaw, Alam, Sansó, Lee, Heitmann, Habib, \& Higdon}]{Holsclaw_2010}
Holsclaw, T., Alam, U., Sansó, B., {et~al.} 2010, Physical Review Letters, 105, \dodoi{10.1103/physrevlett.105.241302}

\bibitem[{Jiménez {et~al.}(2018)Jiménez, Heisenberg, \& Koivisto}]{Jim_nez_2018}
Jiménez, J.~B., Heisenberg, L., \& Koivisto, T. 2018, Physical Review D, 98, \dodoi{10.1103/physrevd.98.044048}

\bibitem[{Jiménez {et~al.}(2020)Jiménez, Heisenberg, Koivisto, \& Pekar}]{Jim_nez_2020}
Jiménez, J.~B., Heisenberg, L., Koivisto, T., \& Pekar, S. 2020, Physical Review D, 101, \dodoi{10.1103/physrevd.101.103507}

\bibitem[{Khyllep {et~al.}(2023)Khyllep, Dutta, Saridakis, \& Yesmakhanova}]{Khyllep_2023}
Khyllep, W., Dutta, J., Saridakis, E.~N., \& Yesmakhanova, K. 2023, Physical Review D, 107, \dodoi{10.1103/physrevd.107.044022}

\bibitem[{Khyllep {et~al.}(2021)Khyllep, Paliathanasis, \& Dutta}]{Khyllep_2021}
Khyllep, W., Paliathanasis, A., \& Dutta, J. 2021, Physical Review D, 103, \dodoi{10.1103/physrevd.103.103521}

\bibitem[{Lazkoz {et~al.}(2019)Lazkoz, Lobo, Ortiz-Baños, \& Salzano}]{Lazkoz_2019}
Lazkoz, R., Lobo, F.~S., Ortiz-Baños, M., \& Salzano, V. 2019, Physical Review D, 100, \dodoi{10.1103/physrevd.100.104027}

\bibitem[{Li \& Shafieloo(2019)}]{Li_2019}
Li, X., \& Shafieloo, A. 2019, The Astrophysical Journal Letters, 883, L3, \dodoi{10.3847/2041-8213/ab3e09}

\bibitem[{Li {et~al.}(2019)Li, Shafieloo, Sahni, \& Starobinsky}]{Li_2019_2}
Li, X., Shafieloo, A., Sahni, V., \& Starobinsky, A.~A. 2019, The Astrophysical Journal, 887, 153, \dodoi{10.3847/1538-4357/ab535d}

\bibitem[{Li {et~al.}(2016)Li, Gonzalez, Yu, Zhu, \& Alcaniz}]{Li_2016}
Li, Z., Gonzalez, J., Yu, H., Zhu, Z.-H., \& Alcaniz, J. 2016, Physical Review D, 93, \dodoi{10.1103/physrevd.93.043014}

\bibitem[{Lin \& Zhai(2021)}]{Lin_2021}
Lin, R.-H., \& Zhai, X.-H. 2021, Physical Review D, 103, \dodoi{10.1103/physrevd.103.124001}

\bibitem[{Mandal {et~al.}(2020{\natexlab{a}})Mandal, Sahoo, \& Santos}]{Mandal_2020}
Mandal, S., Sahoo, P., \& Santos, J. 2020{\natexlab{a}}, Physical Review D, 102, \dodoi{10.1103/physrevd.102.024057}

\bibitem[{Mandal {et~al.}(2020{\natexlab{b}})Mandal, Wang, \& Sahoo}]{Mandal_2020_2}
Mandal, S., Wang, D., \& Sahoo, P. 2020{\natexlab{b}}, Physical Review D, 102, \dodoi{10.1103/physrevd.102.124029}

\bibitem[{Mehrabi \& Rezaei(2021)}]{Mehrabi_2021}
Mehrabi, A., \& Rezaei, M. 2021, The Astrophysical Journal, 923, 274, \dodoi{10.3847/1538-4357/ac2fff}

\bibitem[{Melia \& Yennapureddy(2018)}]{Melia_2018}
Melia, F., \& Yennapureddy, M.~K. 2018, Journal of Cosmology and Astroparticle Physics, 2018, 034–034, \dodoi{10.1088/1475-7516/2018/02/034}

\bibitem[{Moresco(2015)}]{Moresco_2015}
Moresco, M. 2015, Monthly Notices of the Royal Astronomical Society: Letters, 450, L16–L20, \dodoi{10.1093/mnrasl/slv037}

\bibitem[{Moresco {et~al.}(2012)Moresco, Cimatti, Jimenez, Pozzetti, Zamorani, Bolzonella, Dunlop, Lamareille, Mignoli, Pearce, Rosati, Stern, Verde, Zucca, Carollo, Contini, Kneib, Fèvre, Lilly, Mainieri, Renzini, Scodeggio, Balestra, Gobat, McLure, Bardelli, Bongiorno, Caputi, Cucciati, de~la Torre, de~Ravel, Franzetti, Garilli, Iovino, Kampczyk, Knobel, Kovač, Borgne, Brun, Maier, Pelló, Peng, Perez-Montero, Presotto, Silverman, Tanaka, Tasca, Tresse, Vergani, Almaini, Barnes, Bordoloi, Bradshaw, Cappi, Chuter, Cirasuolo, Coppa, Diener, Foucaud, Hartley, Kamionkowski, Koekemoer, López-Sanjuan, McCracken, Nair, Oesch, Stanford, \& Welikala}]{M_Moresco_2012}
Moresco, M., Cimatti, A., Jimenez, R., {et~al.} 2012, Journal of Cosmology and Astroparticle Physics, 2012, 006–006, \dodoi{10.1088/1475-7516/2012/08/006}

\bibitem[{Moresco {et~al.}(2016)Moresco, Pozzetti, Cimatti, Jimenez, Maraston, Verde, Thomas, Citro, Tojeiro, \& Wilkinson}]{Moresco_2016}
Moresco, M., Pozzetti, L., Cimatti, A., {et~al.} 2016, Journal of Cosmology and Astroparticle Physics, 2016, 014–014, \dodoi{10.1088/1475-7516/2016/05/014}

\bibitem[{Nojiri \& Odintsov(2011)}]{Nojiri_2011}
Nojiri, S., \& Odintsov, S.~D. 2011, Physics Reports, 505, 59–144, \dodoi{10.1016/j.physrep.2011.04.001}

\bibitem[{Oka {et~al.}(2014)Oka, Saito, Nishimichi, Taruya, \& Yamamoto}]{Oka_2014}
Oka, A., Saito, S., Nishimichi, T., Taruya, A., \& Yamamoto, K. 2014, Monthly Notices of the Royal Astronomical Society, 439, 2515–2530, \dodoi{10.1093/mnras/stu111}

\bibitem[{Peebles \& Ratra(2003)}]{Peebles_2003}
Peebles, P. J.~E., \& Ratra, B. 2003, Reviews of Modern Physics, 75, 559–606, \dodoi{10.1103/revmodphys.75.559}

\bibitem[{Pinho {et~al.}(2018)Pinho, Casas, \& Amendola}]{Pinho_2018}
Pinho, A.~M., Casas, S., \& Amendola, L. 2018, Journal of Cosmology and Astroparticle Physics, 2018, 027–027, \dodoi{10.1088/1475-7516/2018/11/027}

\bibitem[{Rasmussen \& Williams(2005)}]{exp1}
Rasmussen, C.~E., \& Williams, C. K.~I. 2005, {Gaussian Processes for Machine Learning}, 1st edn. (The MIT Press).
\newblock \url{https://doi.org/10.7551/mitpress/3206.001.0001}

\bibitem[{Ratsimbazafy {et~al.}(2017)Ratsimbazafy, Loubser, Crawford, Cress, Bassett, Nichol, \& Väisänen}]{Ratsimbazafy_2017}
Ratsimbazafy, A.~L., Loubser, S.~I., Crawford, S.~M., {et~al.} 2017, Monthly Notices of the Royal Astronomical Society, 467, 3239–3254, \dodoi{10.1093/mnras/stx301}

\bibitem[{Rau {et~al.}(2019)Rau, Wilson, \& Mandelbaum}]{Rau_2019}
Rau, M.~M., Wilson, S., \& Mandelbaum, R. 2019, Monthly Notices of the Royal Astronomical Society, 491, 4768–4782, \dodoi{10.1093/mnras/stz3295}

\bibitem[{Ren {et~al.}(2022)Ren, Yan, Zhao, Cai, \& Saridakis}]{Ren_2022N}
Ren, X., Yan, S.-F., Zhao, Y., Cai, Y.-F., \& Saridakis, E.~N. 2022, The Astrophysical Journal, 932, 131, \dodoi{10.3847/1538-4357/ac6ba5}

\bibitem[{Seikel \& Clarkson(2013)}]{exp2}
Seikel, M., \& Clarkson, C. 2013, Optimising Gaussian processes for reconstructing dark energy dynamics from supernovae.
\newblock \doarXiv{1311.6678}

\bibitem[{Seikel {et~al.}(2012)Seikel, Clarkson, \& Smith}]{Seikel_2012}
Seikel, M., Clarkson, C., \& Smith, M. 2012, Journal of Cosmology and Astroparticle Physics, 2012, 036–036, \dodoi{10.1088/1475-7516/2012/06/036}

\bibitem[{Simon {et~al.}(2005)Simon, Verde, \& Jimenez}]{Simon_2005}
Simon, J., Verde, L., \& Jimenez, R. 2005, Physical Review D, 71, \dodoi{10.1103/physrevd.71.123001}

\bibitem[{Sokoliuk {et~al.}(2023)Sokoliuk, Arora, Praharaj, Baransky, \& Sahoo}]{Sokoliuk_2023}
Sokoliuk, O., Arora, S., Praharaj, S., Baransky, A., \& Sahoo, P.~K. 2023, Monthly Notices of the Royal Astronomical Society, 522, 252–267, \dodoi{10.1093/mnras/stad968}

\bibitem[{Stern {et~al.}(2010)Stern, Jimenez, Verde, Kamionkowski, \& Stanford}]{Daniel_2010}
Stern, D., Jimenez, R., Verde, L., Kamionkowski, M., \& Stanford, S.~A. 2010, Journal of Cosmology and Astroparticle Physics, 2010, 008–008, \dodoi{10.1088/1475-7516/2010/02/008}

\bibitem[{Vagnozzi(2020)}]{Vagnozzi_2020}
Vagnozzi, S. 2020, Physical Review D, 102, \dodoi{10.1103/physrevd.102.023518}

\bibitem[{Verde {et~al.}(2014)Verde, Protopapas, \& Jimenez}]{verde2014expansion}
Verde, L., Protopapas, P., \& Jimenez, R. 2014, The expansion rate of the intermediate Universe in light of Planck.
\newblock \doarXiv{1403.2181}

\bibitem[{Wang \& Meng(2016)}]{wang2016modelindependent}
Wang, D., \& Meng, X.-H. 2016, Model-independent determination on $H_0$ using the latest $H(z)$ data.
\newblock \doarXiv{1610.01202}

\bibitem[{Wang {et~al.}(2017)Wang, Zhao, Chuang, Ross, Percival, Gil-Marín, Cuesta, Kitaura, Rodriguez-Torres, Brownstein, Eisenstein, Ho, Kneib, Olmstead, Prada, Rossi, Sánchez, Salazar-Albornoz, Thomas, Tinker, Tojeiro, Vargas-Magaña, \& Zhu}]{Wang_2017}
Wang, Y., Zhao, G.-B., Chuang, C.-H., {et~al.} 2017, Monthly Notices of the Royal Astronomical Society, 469, 3762–3774, \dodoi{10.1093/mnras/stx1090}

\bibitem[{Wong {et~al.}(2019)Wong, Suyu, Chen, Rusu, Millon, Sluse, Bonvin, Fassnacht, Taubenberger, Auger, Birrer, Chan, Courbin, Hilbert, Tihhonova, Treu, Agnello, Ding, Jee, Komatsu, Shajib, Sonnenfeld, Blandford, Koopmans, Marshall, \& Meylan}]{Wong_2019}
Wong, K.~C., Suyu, S.~H., Chen, G. C.-F., {et~al.} 2019, Monthly Notices of the Royal Astronomical Society, 498, 1420–1439, \dodoi{10.1093/mnras/stz3094}

\bibitem[{Xu {et~al.}(2019)Xu, Li, Harko, \& Liang}]{Xu_2019}
Xu, Y., Li, G., Harko, T., \& Liang, S.-D. 2019, The European Physical Journal C, 79, \dodoi{10.1140/epjc/s10052-019-7207-4}

\bibitem[{Yin \& Wei(2019)}]{Yin_2019}
Yin, Z.Y., and Wei, H. 2019, Science China Physics, Mechanics and Astronomy, 62, \dodoi{10.1007/s11433-019-9373-0}

\bibitem[{Zhang \& Li(2018)}]{Zhang_2018}
Zhang, M.-J., \& Li, H. 2018, The European Physical Journal C, 78, \dodoi{10.1140/epjc/s10052-018-5953-3}

\end{thebibliography}
\bibliographystyle{aasjournal}

\end{document}